\documentclass[pra,twocolumn]{revtex4-2}
\usepackage{enumitem}
\usepackage{amsfonts}
\usepackage{mathrsfs}
\usepackage{graphicx}
\usepackage{amssymb}
\usepackage{pifont}
\usepackage{float}
\usepackage{amsmath}
\usepackage{hyperref}

\usepackage{epstopdf}
\usepackage{cleveref}
\usepackage{amsthm}
\usepackage{physics}
\usepackage{bm}
\usepackage{booktabs}

\hypersetup{
    colorlinks=true,
    linkcolor=blue,
    citecolor=blue,
    urlcolor=blue
}

\def\cH{\mathcal{H}}

\def\cB{\mathcal{B}}
\def\D{\mathcal{D}}

\def\cI{\mathcal{I}}
\def\cR{\mathcal{R}}

\def\J{\mathcal{J}}

\def\F{\mathcal{F}}

\newcommand{\cC}{\mathcal{C}}

\newtheorem{dfn}{Definition}
\newtheorem{thm}{Theorem}
\newtheorem{prop}{Proposition}

\newtheorem{eg}{Example}

\begin{document}
\title{Resource theory of \(\pi\)-ordered imaginarity and its resource quantifiers}
	\author{Yanjun Chu}
	\email{chuyj@henu.edu.cn}
\author{Yushan Ding}
	\author{Yuhang Xie}
\affiliation{School of Mathematics and Statistics, Henan University, Kaifeng, 475004, China}
\author{Chenyang Cui}
\affiliation{School of Mathematics and Statistics, Henan University, Kaifeng, 475004, China}
\author{Shao-Ming Fei}
\email{feishm@cnu.edu.cn}
\affiliation{School of Mathematics Sciences, Capital Normal University, Beijing 100048, China}

\begin{abstract}
Conventional resource theory of quantum imaginarity takes real density matrices as free states without ordering constraints on their diagonal populations. Motivated by the role of population ordering in the incoherent ergotropy of quantum batteries and the recently proposed framework of \(\pi\)-ordered coherence, we develop the resource theory of \(\pi\)-ordered imaginarity. In this refined framework, free states are required to be real in the reference basis and to obey a prescribed permutation order \(\pi\) for their diagonal populations. This theory thereby identifies two independent quantum resources: complex phases and population misordering, where the latter is inaccessible within the standard imaginarity resource theory. We characterize the structure of free states using majorization and the Schur–Horn theorem, define admissible \(\pi\)-ordered real operations, and examine their intersection with \(\pi\)-ordered incoherent operations. We further construct resource quantifiers for the \(\pi\)-ordered imaginarity based on relative entropy, fidelity and robustness, and derive inequalities relating the robustness of \(\pi\)-ordered imaginarity and that of \(\pi\)-ordered coherence. The present framework generalizes the standard resource theory of imaginarity to incorporate population ordering and is naturally suited for thermodynamic settings such as quantum batteries.
\end{abstract}

	\maketitle

\section{Introduction}
Quantum imaginarity constitutes a fundamental physical resource enabling distinctive capabilities in modern quantum technologies. Hickey and Gour initiated the resource theory of quantum imaginarity to characterize the operational role of complex-valued structures in quantum information processing \cite{Hickey2018}. This framework provides a rigorous scheme in quantifying imaginarity and has laid the foundation for treating complex components of quantum states as a genuine quantum resource \cite{Chitambar2019}.

A central problem in this resource theory is the quantitative characterization of imaginarity. Given a fixed reference basis \(\{|i\rangle\}_{i=1}^{d}\), free states are defined as density matrices with purely real entries. States that cannot be brought into real form carry imaginarity resource \cite{Wu2021}. The theory has revealed the operational relevance of complex phases across tasks including quantum state manipulation, state discrimination, and optical implementations of quantum protocols \cite{Wu2021,Wu2021prl,Wu2024}.

A rich family of imaginarity quantifiers with clear operational interpretations has been developed, such as the robustness of imaginarity \cite{Hickey2018,Wu2021}, fidelity-based imaginarity \cite{Wu2021prl}, and geometric imaginarity \cite{Wu2021}, alongside many other related measures \cite{XueQIP383,chen220311,XuPLA130024,ZhangQIP361,chen280312,LiuJPA035302,WuCTP095101,TianPLA130479,ZhangEPJP669,DuPRA022405}. These quantifiers help build physical intuition about imaginarity and highlight advantages of complex quantum states over real ones. Using state-channel duality, the notion of imaginarity has also been extended from states to quantum channels, yielding a resource theory for channel imaginarity \cite{ChenPLA130129,Wu230316,WangEPJP216,FuQIP140}.

Recent studies on the ergotropy of quantum batteries have underscored the importance of population ordering among diagonal elements \cite{Wang2026}. Nevertheless, standard imaginarity resource theory only distinguishes real and nonreal density matrices while ignoring ordering constraints on diagonal populations. Inspired by the recently proposed \(\pi\)-ordered coherence theory, in which a fixed permutation order is imposed on diagonal entries of incoherent free states to produce a resource structure different from standard coherence \cite{Baumgratz2014,Qi2026}, we introduce a refined framework enforcing two simultaneous constraints: reality and population ordering. In this \(\pi\)-ordered imaginarity theory, a state is resourceful if it contains nonremovable imaginary components or if its diagonal populations violate the prescribed permutation order \(\pi\). The theory thus captures two independent resources, complex phases and population misordering, the latter of which is captured within the conventional imaginarity.

Population redistribution and inversion are central to the charging dynamics of quantum batteries. It has been shown that the ergotropy equals coherent ergotropy in the absence of population inversion. When local or global inversion occurs, both coherent and incoherent contributions enter the ergotropy \cite{Wang2026}. Consider a quantum battery described by a fixed Hamiltonian
\(H=\sum_{k=1}^d \varepsilon_k |\varepsilon_k\rangle\langle\varepsilon_k|\),  \(0\le\varepsilon_1\le \varepsilon_2\le\cdots\le \varepsilon_d,\) with energy eigenbasis \(\{|\varepsilon_k\rangle\}_{k=1}^d\). The average energy of a battery state \(\rho\) is \(E(\rho)=\operatorname{tr}[H\rho]\). We recall the incoherent contribution to ergotropy \cite{Francica2020}. Let \(\pi\) denote the permutation arranging the populations of \(\rho\) in nonincreasing order, \(\rho_{\pi_j,\pi_j} \ge \rho_{\pi_{j+1},\pi_{j+1}}\) for all \(j=1,\dots,d-1\). Define \(\sigma_\rho = P_{\pi} \rho P_{\pi}^\dagger = \sum_{k}\sum_{k'} \rho_{\pi_k,\pi_{k'}} |\varepsilon_k\rangle\langle\varepsilon_{k'}|\), where \(P_{\pi}\) is the permutation matrix associated with \(\pi\); \(\sigma_\rho\) minimizes the average energy. The incoherent ergotropy reads
\(\mathcal{E}_i(\rho) = \operatorname{Tr}\bigl\{ H(\rho - \sigma_\rho) \bigr\} = \sum_k \varepsilon_k \big(\rho_{kk} - \rho_{\pi_k \pi_k}\big),\)
where \(\rho_{kk} = \langle \varepsilon_k | \rho | \varepsilon_k \rangle\) is the population of \(\rho\) on the k-th energy eigenstate. Since incoherent ergotropy depends solely on energy populations, it shapes the dynamics of total ergotropy. This observation provides our primary motivation for developing \(\pi\)-ordered imaginarity.

The free-state structure we adopt has a natural connection to majorization and the Schur–Horn theorem. For a real state \(\rho\), define \(\widetilde\rho=P_\pi\rho P_\pi^\dagger\). Because \(\rho\) is real and Hermitian, \(\widetilde\rho\) is real symmetric. By the spectral theorem, there exists an orthogonal matrix \(O_\rho\in O(d)\) such that
\(\widetilde\rho=O_\rho\,\operatorname{diag}(\lambda_1^\downarrow,\ldots,\lambda_d^\downarrow)O_\rho^{\mathsf T},\)
with \(\lambda_1^\downarrow\ge\lambda_2^\downarrow\ge\cdots\ge \lambda_d^\downarrow\ge0\). The diagonal entries of \(\widetilde\rho\) are
\(p_i=\widetilde\rho_{ii} =\sum_{j=1}^{d}O_{ij}^{\,2}\lambda_j .\)
Let \(B_{ij}=O_{ij}^{\,2}\). The matrix B is doubly stochastic, satisfying \(\boldsymbol{p}=B\boldsymbol{\lambda}\), where \(\boldsymbol{p}=(p_1, p_2,\dots, p_d)\) and \(\boldsymbol{\lambda}=(\lambda_1,\lambda_2,\dots,\lambda_d)\). A vector \(\boldsymbol y\) is majorized by \(\boldsymbol x\), written \(\boldsymbol{y}\prec \boldsymbol{x}\), if and only if \(\boldsymbol y = D\boldsymbol x\) for some doubly stochastic matrix \(D\) \cite{Marshall2011}. It follows that \(\boldsymbol{\lambda}(\rho)\succ \boldsymbol{p}(\rho)\), i.e.,
\(\sum_{i=1}^{k}\lambda_i \ge \sum_{i=1}^{k}p_i, k=1,\ldots,d-1, \sum_{i=1}^{d}\lambda_i = \sum_{i=1}^{d}p_i=1 .\)
The Schur–Horn theorem states that, given the nonincreasingly sorted eigenvalue vector \(\boldsymbol{\lambda}\) of a real symmetric matrix, a vector \(\boldsymbol{p}\) can be its diagonal if and only if \(\boldsymbol{\lambda}\succ \boldsymbol{p}\) \cite{Marshall2011}. Therefore, for fixed sorted spectrum \(\boldsymbol{\lambda}^\downarrow\), admissible diagonal vectors are exactly those majorized by \(\boldsymbol{\lambda}^\downarrow\). The \(\pi\)-order condition selects a permutation-ordered subset within this Schur–Horn polytope. 

We retain vectors \(\boldsymbol{p}\) satisfying the nonincreasing monotonicity condition as the diagonal entries of \(\rho\); the resulting matrices form the set \(\{P_\pi\rho P_\pi^\dagger\}\), whose elements satisfy both reality and population ordering.There exists a unitary mapping relating \(\pi\)-ordered incoherent states and \(\pi\)-ordered real states. Let \(U_\rho=P_\pi^\dagger O_\rho P_\pi\). Then
\(U_\rho^\dagger\rho U_\rho = P_\pi^\dagger \operatorname{diag}(\lambda_1^\downarrow,\ldots,\lambda_d^\downarrow) P_\pi \in\mathcal F_\pi\)
for any \(\rho\in\mathcal R_\pi\). Every \(\pi\)-ordered real state thus admits a diagonal representative in \(\mathcal{F}_\pi\) sharing its full spectrum. In this sense, \(\mathcal{F}_\pi\) can be viewed as a canonical diagonal slice obtained by orthogonally diagonalizing elements of \(\mathcal{R}_\pi\). We emphasize that \(O_\rho\) is generally state-dependent, so this relation describes a geometric correspondence between the two families of states.

In this paper, we construct the resource theory of \(\pi\)-ordered imaginarity, where the set of free states adapts to an arbitrary permutation \(\pi\) on \(\{1,2,\dots,d\}\). The paper is organized as follows. In Sec.~\ref{sect2}, we define \(\pi\)-ordered real states as the free states of the theory. Sec.~\ref{sect3} introduces \(\pi\)-ordered real operations, characterizes their structure, and discusses their intersection with \(\pi\)-ordered incoherent operations. In Sec.~\ref{sect4}, we establish the general definition of \(\pi\)-ordered imaginarity measures and construct specific quantifiers induced by relative entropy, fidelity and robustness. We further investigate the geometric relation between \(\pi\)-ordered imaginarity and \(\pi\)-ordered coherence, including inequalities connecting their robustness measures. Sec.\ref{sec5} concludes the work.

\section{Free states: $\pi$-ordered real states}\label{sect2}
In Ref.~\cite{Qi2026}, the notion of $\pi$-ordered incoherent states was introduced.
For completeness, we first recall the set $\mathcal{F}_\pi$ of all $\pi$-ordered incoherent states, defined as
\begin{equation}
\mathcal{F}_{\pi}
=\{
\sigma
=\sum_{i=1}^{d} \sigma_{i}|i\rangle\langle i|~|\sigma_{i}\geq 0,i=1,\cdots,d
\},
\end{equation}
with constraints
\begin{equation*}
\sigma_{\pi^{-1}(1)} \geq \sigma_{\pi^{-1}(2)} \geq \cdots \geq \sigma_{\pi^{-1}(d)} \geq 0,\qquad
\sum_{i=1}^{d}\sigma_i = 1.
\end{equation*}

Let $\mathcal{H}$ be a finite-dimensional Hilbert space of dimension $d$.
We use $\mathcal{L}(\mathcal{H})$ to denote the space of linear operators on $\mathcal{H}$, $\mathcal{D}(\mathcal{H})$ the set of density operators on $\mathcal{H}$, and $\mathcal{B}(\mathcal{H})$ the set of bounded linear operators on $\mathcal{H}$.
In this section, we introduce the definition of $\pi$-ordered real states.

Let $\{\ket{i}\}_{i=1}^d$ be a fixed orthonormal basis of $\mathcal{H}$, and $\pi=(\pi(1),\pi(2),\dots,\pi(d))$  an arbitrary permutation of $\{1,2,\dots,d\}$. Denote $P_\pi$ the associated permutation matrix satisfying $P_\pi\ket{i}=\ket{\pi(i)}$ for all $i$. A density matrix $\rho\in\mathcal{D}(\mathcal{H})$ is called a \textit{$\pi$-ordered real state} if
\begin{equation}
	\rho=\sum_{i,j=1}^d \rho_{ij}\ket{i}\bra{j},
\end{equation}
where all matrix entries $\rho_{ij}\in\mathbb{R}$, and
\begin{equation}
	P_\pi\rho P_\pi^\dagger=\sum_{i,j=1}^d \rho_{\pi^{-1}(i)\pi^{-1}(j)}\ket{i}\bra{j}
\end{equation}
with $\rho_{\pi^{-1}(1)\pi^{-1}(1)}\ge \rho_{\pi^{-1}(2)\pi^{-1}(2)}\ge\cdots\ge \rho_{\pi^{-1}(d)\pi^{-1}(d)}.$

States not satisfying this condition are referred to as \textit{$\pi$-ordered imaginary states}.
We write $\mathcal{R}_\pi$ for the collection of all $\pi$-ordered real states:
\begin{equation}
	\mathcal{R}_\pi=\Bigl\{
	\rho\in\mathcal{D}(\mathcal{H}):\;
	\rho=\sum_{i,j=1}^d \rho_{ij}\ket{i}\bra{j},\; \rho_{ij}\in\mathbb{R}\Bigr\},\end{equation}
with constraints
\(\rho_{\pi^{-1}(1)\pi^{-1}(1)}\ge \rho_{\pi^{-1}(2)\pi^{-1}(2)}\ge\cdots\ge \rho_{\pi^{-1}(d)\pi^{-1}(d)}.\)

One can readily check that $\mathcal{F}_\pi\subset \mathcal{R}_\pi$ and that $\mathcal{R}_\pi$ is convex.
Let $\mathcal{R}$ denote the set of all real density matrices. Then the inclusion $\mathcal{R}_\pi\subset \mathcal{R}$ holds.
\section{Free operations: $\pi$-ordered real operations}\label{sect3}
Owing to the distinctive role of imaginary numbers played in quantum theory, Hickey and Gour \cite{Hickey2018} introduced the resource theory of imaginarity.
In that framework, the free states are real states: density matrices whose matrix entries are real with respect to a fixed reference basis.
The free operations, called real operations, are completely positive and trace-preserving (CPTP) maps $\Lambda(\cdot)=\sum_j K_j \cdot K_j^\dagger$ for which every Kraus operator $K_j$ has real matrix elements.

In our resource theory of $\pi$-ordered imaginarity, the free states are the $\pi$-ordered real states defined in the previous section.
Accordingly, the free operations are taken to be CPTP maps that map the set $\mathcal{R}_\pi$ of $\pi$-ordered real states into itself.

\begin{dfn}[$\pi$-ordered real operation]\label{d1}
Suppose $\Phi\colon \mathcal{B}(\mathcal{H})\to\mathcal{B}(\mathcal{H})$ is a quantum operation with Kraus decomposition $\{K_n\}\subset \mathcal{B}(\mathcal{H})$ satisfying $\sum_{n} K_n^\mathsf{T} K_n=\mathbb{I}$ and $\bra{i}K_n\ket{j}\in\mathbb{R}$ for all $i,j\in\{1,\dots,d\}$. We call $\Phi$ a \textit{$\pi$-ordered real operation} ($\pi$-ORO) if $K_n \mathcal{R}_\pi K_n^\mathsf{T} \subseteq \mathbb{R}\mathcal{R}_\pi$ for every $n$.
\end{dfn}

It follows directly from Definition \ref{d1} that every $\pi$-ORO sends $\pi$-ordered real states to $\pi$-ordered real states.
Moreover, any $\pi$-ORO is automatically a real operation.
The following theorem gives an equivalent characterization of $\pi$-OROs.

\begin{thm}\label{thm:pi_order_real_operation}
A quantum operation $\Phi(\cdot)=\sum_n K_n (\cdot) K_n^\mathsf{T}$ is a $\pi$-ORO if and only if both conditions below hold:
\begin{enumerate}
    \item $\sum_n K_n^\mathsf{T} K_n = \mathbb{I}$, and $\langle i|K_n|j\rangle\in\mathbb{R}$ for all $i,j\in\{1,\dots,d\}$;
    \item for all $n$ and $a$, there exist scalars $\lambda_l\ge0$ such that
    \[
    \widetilde K_n^\mathsf{T} A_a \widetilde K_n - \sum_{l=1}^{d-1}\lambda_l A_l \succeq 0,
    \]
\end{enumerate}
where $A_l=\ket{l}\bra{l}-\ket{l+1}\bra{l+1}$ for $1\le l\le d-1$, $\widetilde K_n=P_\pi K_n P_\pi^\mathsf{T}$, and $X\succeq0$ means that $X$ is positive semidefinite.
\end{thm}
A detailed proof of this theorem can be found in Appendix \ref{app:proof of theorem 2}.

In general, real Kraus operators need not map a basis states to certain given basis states.
Instead, the off-diagonal entries of the output are jointly controlled by the row vectors of $\widetilde K_n$.
We illustrate the preceding characterization with an explicit example.

\begin{eg}\label{ex:3}
Take the permutation $\pi(1,2,3)=(3,1,2)$ and $d=3$. Let $\Phi$ be a quantum operation with Kraus operators $\{K_1,K_2,K_3\}$ of the form $K_n=u v_n^\mathsf{T}$, where
\[
u=\frac{1}{\sqrt{14}}\begin{pmatrix}1\\3\\2\end{pmatrix},
v_1=\frac13\begin{pmatrix}1\\2\\2\end{pmatrix},
v_2=\frac13\begin{pmatrix}2\\1\\-2\end{pmatrix},
v_3=\frac13\begin{pmatrix}-2\\2\\-1\end{pmatrix}.
\]
Then
\[
K_1=\frac{1}{3\sqrt{14}}
\begin{pmatrix}
	1 & 2 & 2\\
	3 & 6 & 6\\
	2 & 4 & 4
\end{pmatrix},
K_2=\frac{1}{3\sqrt{14}}
\begin{pmatrix}
	2 & 1 & -2\\
	6 & 3 & -6\\
	4 & 2 & -4
\end{pmatrix},\]
\[
K_3=\frac{1}{3\sqrt{14}}
\begin{pmatrix}
	-2 & 2 & -1\\
	-6 & 6 & -3\\
	-4 & 4 & -2
\end{pmatrix}.
\]
All $K_n$ are real and satisfy $\sum_n K_n^\mathsf{T}K_n=\mathbb{I}$.

For $\pi(1,2,3)=(3,1,2)$, the permutation matrix is
\[
P_\pi=
\begin{pmatrix}
	0 & 1 & 0\\
	0 & 0 & 1\\
	1 & 0 & 0
\end{pmatrix}.
\]
Define $u'=P_\pi u=\frac{1}{\sqrt{14}}\begin{pmatrix}3\\2\\1\end{pmatrix}$ and $\omega_n=P_\pi v_n$. Then
$\widetilde K_n=P_\pi K_n P_\pi^\mathsf{T}=P_\pi u v_n^\mathsf{T}P_\pi^\mathsf{T}=u' \omega_n^\mathsf{T}$
and $\widetilde K_n^\mathsf{T} A_a \widetilde K_n
=\omega_n (u')^\mathsf{T} A_a u' \omega_n^\mathsf{T}$.
Recall $A_1=\operatorname{diag}(1,-1,0)$ and $A_2=\operatorname{diag}(0,1,-1)$.
Direct computation gives
\[
(u')^\mathsf{T}A_1 u'=\frac{9-4}{14}=\frac{5}{14},\qquad
(u')^\mathsf{T}A_2 u'=\frac{4-1}{14}=\frac{3}{14}.
\]
Therefore
\[
\widetilde K_n^\mathsf{T}A_1\widetilde K_n=\frac{5}{14}\,\omega_n \omega_n^\mathsf{T}\succeq 0,\quad
\widetilde K_n^\mathsf{T}A_2\widetilde K_n=\frac{3}{14}\,\omega_n \omega_n^\mathsf{T}\succeq 0.
\]
For all $n=1,2,3$ and $a=1,2$, we may take $\lambda_1=\lambda_2=0$.
The conditions of Theorem~\ref{thm:pi_order_real_operation} are satisfied, and $\Phi$ is a $\pi$-ORO.
\end{eg}

Example~\ref{ex:3} is a simple case in which each $\widetilde K_n^\mathsf{T}A_a\widetilde K_n$ is already positive semidefinite, so we can choose all $\lambda_l=0$.
We next give an example where nontrivial $\lambda_l>0$ are required for $n=1$.

\begin{eg}
Let $d=3$, $\pi(1,2,3)=(3,1,2)$, and $K_n=u v_n^\mathsf{T}$, where
\[
u=\frac1{\sqrt{14}} \begin{pmatrix} 1\\3\\2 \end{pmatrix},\quad
v_1=\frac1{\sqrt{11}}\begin{pmatrix}1\\3\\1\end{pmatrix},\]
\[
v_2=\frac1{\sqrt{18}}\begin{pmatrix}-4\\1\\1\end{pmatrix},\quad
v_3=\frac1{\sqrt{198}} \begin{pmatrix} 2\\-5\\13 \end{pmatrix}.
\]
Then
\[
K_1=\frac1{\sqrt{154}} \begin{pmatrix} 1&3&1\\ 3&9&3\\ 2&6&2 \end{pmatrix},
\quad
K_2=\frac1{\sqrt{252}} \begin{pmatrix} -4&1&1\\ -12&3&3\\ -8&2&2 \end{pmatrix},
\]
\[
K_3=\frac1{\sqrt{2772}} \begin{pmatrix} 2&-5&13\\ 6&-15&39\\ 4&-10&26 \end{pmatrix}.
\]
All $K_n$ are real and satisfy $\sum_n K_n^\mathsf{T}K_n=\mathbb{I}$.

With respect to $\pi(1,2,3)=(3,1,2)$, one has
\[
P_\pi=
\begin{pmatrix}
	0 & 1 & 0\\
	0 & 0 & 1\\
	1 & 0 & 0
\end{pmatrix}.
\]
Set $u'=P_\pi u=\frac{1}{\sqrt{14}}\begin{pmatrix}3\\2\\1\end{pmatrix}$ and $\omega_n=P_\pi v_n$. Then
$\widetilde K_n=P_\pi K_n P_\pi^\mathsf{T}=u' \omega_n^\mathsf{T}$ and
$\widetilde K_n^\mathsf{T} A_a \widetilde K_n
=\omega_n (u')^\mathsf{T} A_a u' \omega_n^\mathsf{T}$.
With $A_1=\operatorname{diag}(1,-1,0)$ and $A_2=\operatorname{diag}(0,1,-1)$,
we still have
\[
(u')^\mathsf{T}A_1 u'=\frac{5}{14},\quad
(u')^\mathsf{T}A_2 u'=\frac{3}{14}.
\]
Hence,
\[
\widetilde K_n^\mathsf{T}A_1\widetilde K_n=\frac{5}{14}\,\omega_n \omega_n^\mathsf{T}\succeq 0,\quad
\widetilde K_n^\mathsf{T}A_2\widetilde K_n=\frac{3}{14}\,\omega_n \omega_n^\mathsf{T}\succeq 0.
\]
Explicitly, for $n=1$,
\[
\widetilde K_1^{\mathsf T}A_1\widetilde K_1 = \frac5{154} \begin{pmatrix} 9 & 3 & 3 \\ 3 & 1 & 1 \\ 3 & 1 & 1 \end{pmatrix},\quad
\widetilde K_1^{\mathsf T}A_2\widetilde K_1 = \frac3{154} \begin{pmatrix} 9 & 3 & 3 \\ 3 & 1 & 1 \\ 3 & 1 & 1 \end{pmatrix}.
\]
For $n=2$,
\[
\widetilde K_2^{\mathsf T}A_1\widetilde K_2 = \frac5{252} \begin{pmatrix} 1 & 1 & -4 \\ 1 & 1 & -4 \\ -4 & -4 & 16 \end{pmatrix},\]\[
\widetilde K_2^{\mathsf T}A_2\widetilde K_2 = \frac1{84} \begin{pmatrix} 1 & 1 & -4 \\ 1 & 1 & -4 \\ -4 & -4 & 16 \end{pmatrix}.
\]
For $n=3$,
\[
\widetilde K_3^{\mathsf T}A_1\widetilde K_3 = \frac5{2772} \begin{pmatrix} 25 & -65 & -10 \\ -65 & 169 & 26 \\ -10 & 26 & 4 \end{pmatrix},\]
\[
\widetilde K_3^{\mathsf T}A_2\widetilde K_3 = \frac1{924} \begin{pmatrix} 25 & -65 & -10 \\ -65 & 169 & 26 \\ -10 & 26 & 4 \end{pmatrix}.
\]

Choose $\lambda_1=1/{1000}$ and $\lambda_2=1/{2000}$ for the two constraints $a=1,2$ associated with $n=1$. Then
\[
\lambda_1A_1+\lambda_2A_2 = \operatorname{diag} \left( \frac1{1000},-\frac1{2000},-\frac1{2000} \right).
\]
Define $M(q)= q\omega_1\omega_1^{\mathsf T} -\lambda_1A_1-\lambda_2A_2$, where $q=5/{14}$ or $q=3/{14}$.
All the principal minors
\[
M_{11}=\frac{9q}{11}-\frac1{1000}>0,\quad
M_{22}=M_{33}=\frac q{11}+\frac1{2000}>0,
\]
\[
\Delta_{12}=\Delta_{13} = \frac1{1000} \left( \frac{7q}{22}-\frac1{2000} \right)>0,\]
\[\Delta_{23} = \frac q{11000}+\frac1{16\,000\,000}>0,
\]
and the determinant
\[
\det M(q) = \frac1{16\,000\,000} \left( \frac{5q}{11}-\frac1{1000} \right)>0
\]
for both values of $q$. Therefore,
\[
\widetilde K_1^{\mathsf T}A_a\widetilde K_1 -\frac1{1000}A_1-\frac1{2000}A_2 \succeq0, \quad a=1,2.
\]
For $n=2,3$, take $\lambda_1=\lambda_2=0$. The conditions in Theorem~\ref{thm:pi_order_real_operation} hold for all $n=1,2,3$ and $a=1,2$. Therefore, $\Phi$ is a $\pi$-ORO.
\end{eg}
Having established the characterization of $\pi$-OROs, we now discuss the relation between $\pi$-OROs and $\pi$-ordered incoherent operations ($\pi$-OIOs). Recall the inclusion $\mathcal{F}_\pi\subset \mathcal{R}_\pi$: every $\pi$-ordered incoherent state is a $\pi$-ordered real state.
By definition, a $\pi$-OIO preserves $\mathcal{F}_\pi$, while a $\pi$-ORO preserves $\mathcal{R}_\pi$.
In general, neither class contains the other: a $\pi$-ORO need not be a $\pi$-OIO, and a $\pi$-OIO need not be a $\pi$-ORO.
We combine the Kraus conditions for $\pi$-OROs from Theorem~\ref{thm:pi_order_real_operation} with the known characterization of $\pi$-OIOs in Ref.~\cite{Qi2026} to describe their intersection, i.e., channels that are simultaneously $\pi$-ORO and $\pi$-OIO.

This naturally motivates two questions.
First, given a $\pi$-ORO, what necessary and sufficient conditions guarantee that it maps every $\pi$-ordered incoherent state to a $\pi$-ordered incoherent state. Second, given a $\pi$-OIO, what extra constraints on its Kraus operators ensure that it is also a $\pi$-ORO. We address the first question in the following theorem.
\begin{thm}\label{thm:pi-ORO and pi-OIO-1}
Let $\Phi$ be a $\pi$-ORO with Kraus representation $\Phi(\rho)=\sum_n K_n\rho K_n^\mathsf{T}$, where each $K_n\in\mathbb{R}^{d\times d}$ and $\sum_n K_n^\mathsf{T} K_n=\mathbb{I}$.
Then $\Phi(\mathcal{F}_\pi)\subseteq\mathcal{F}_\pi$
if and only if for $\forall r \neq s$, 
\begin{equation}\label{eq:condC}
\sum_n K_{n,\pi^{-1}(r),\pi^{-1}(j)} K_{n,\pi^{-1}(s),\pi^{-1}(j)} = 0,\quad
\forall j\in\{1,\dots,d\}. \tag{C}
\end{equation}
\end{thm}

\begin{proof}
$(\Rightarrow)$: Assume $\Phi(\mathcal{F}_\pi)\subseteq\mathcal{F}_\pi$.
Define the channel conjugated by the permutation:
\[
\widetilde{\Phi}(X):=P_\pi\Phi(P_\pi^\mathsf{T} X P_\pi)P_\pi^\mathsf{T}
=\sum_n \widetilde K_n X \widetilde K_n^\mathsf{T},
\]
where $\widetilde K_n = P_\pi K_n P_\pi^\mathsf{T}$.
Then $\widetilde{\Phi}$ must send every vertex of $P_\pi\mathcal{F}_\pi P_\pi^\dagger$ to a diagonal matrix.
Take the vertex states
\[
w_l=\frac1l\sum_{j=1}^l \ket{j}\bra{j},\quad l=1,\dots,d.
\]
The $(r,s)$ matrix element of $\widetilde{\Phi}(w_l)$ is
\[
\langle r|\widetilde{\Phi}(w_l)|s\rangle
=\Big\langle r\Big|\sum_n \widetilde K_n w_l \widetilde K_n^\mathsf{T}\Big|s\Big\rangle
=\frac1l\sum_{j=1}^l\sum_n \langle r|\widetilde K_n|j\rangle\langle j|\widetilde K_n^\mathsf{T}|s\rangle.
\]
Set $a_{n,r j}=\langle r|\widetilde K_n|j\rangle=\langle q_r|K_n|q_j\rangle$, where $q_r=\pi^{-1}(r)$. Then
\[
\langle r|\widetilde{\Phi}(w_l)|s\rangle
=\frac1l\sum_{j=1}^l\sum_n a_{n,rj}\, a_{n,sj}.
\]
For $r\neq s$, this must vanish.
Define $C_j^{(r,s)}=\sum_n a_{n,r j}\, a_{n,sj}$.
The condition $\sum_{j=1}^l C_j^{(r,s)}=0$ for all $l=1,\dots,d$ forces $C_j^{(r,s)}=0$ for every $j$, which is exactly \eqref{eq:condC}.

$(\Leftarrow)$: Suppose condition \eqref{eq:condC} holds.
Take any $\sigma\in\mathcal{F}_\pi$ and set
\[
\widetilde\sigma=P_\pi\sigma P_\pi^\mathsf{T}
=\sum_{j=1}^d \beta_j\,\ket{j}\bra{j},
\]
with $\beta_1\ge\beta_2\ge\cdots\ge\beta_d\ge0$ and $\sum_j\beta_j=1$.
The off-diagonal entries of $\widetilde{\Phi}(\widetilde\sigma)$ are
\[
\langle r|\widetilde{\Phi}(\widetilde\sigma)|s\rangle
=\sum_{n,j}\beta_j\,\langle r|\widetilde K_n|j\rangle\langle j|\widetilde K_n^\mathsf{T}|s\rangle
=\sum_{j}\beta_j C_j^{(r,s)}.
\]
If $r\neq s$, condition \eqref{eq:condC} gives $\langle r|\widetilde{\Phi}(\widetilde\sigma)|s\rangle=0$, so the output is diagonal.
Since $\Phi$ is a $\pi$-ORO and $\sigma\in\mathcal{F}_\pi\subseteq\mathcal{R}_\pi$, we have $\Phi(\sigma)\in\mathcal{R}_\pi$, which enforces the required $\pi$-ordering of its diagonal entries.
Hence $\Phi(\sigma)\in\mathcal{F}_\pi$ and $\Phi(\mathcal{F}_\pi)\subseteq\mathcal{F}_\pi$.
\end{proof}

We illustrate this result with an example.
\begin{eg}\label{ex:pi-ORO and pi-OIO-1}
Let $d=3$, $\pi(1,2,3)=(3,1,2)$, and let $\Phi$ have Kraus operators $\{K_1,K_2,K_3\}$.
Take the orthogonal matrix
\[
R=\frac13 \begin{pmatrix} 2&-1&2\\ 2&2&-1\\ -1&2&2 \end{pmatrix}, \quad R^\mathsf{T}R=I_3,
\]
and let $r_n^\mathsf{T}$ denote its $n$-th row. Define $K_n=\ket{2}\bra{r_n|}$. Then
\[
K_1=\frac13 \begin{pmatrix} 0&0&0\\ 2&-1&2\\ 0&0&0 \end{pmatrix}, \quad K_2=\frac13 \begin{pmatrix} 0&0&0\\ 2&2&-1\\ 0&0&0 \end{pmatrix}, \]
\[ K_3=\frac13 \begin{pmatrix} 0&0&0\\ -1&2&2\\ 0&0&0 \end{pmatrix}.
\]
All $K_n$ are real and satisfy $\sum_n K_n^\mathsf{T}K_n=\mathbb{I}$. Let
\[
A_1=\operatorname{diag}(1,-1,0),\quad A_2=\operatorname{diag}(0,1,-1).
\]
Because $P_\pi\ket{2}=\ket{1}$, we have $\widetilde K_n =P_\pi K_nP_\pi^\mathsf{T} =\ket{1}\bra{\omega_n}$ with $\omega_n=P_\pi r_n$.
Thus
\[
\widetilde K_n^\mathsf{T}A_a\widetilde K_n = \langle1|A_a|1\rangle\,\omega_n\omega_n^\mathsf{T}.
\]
Moreover,
\[
\langle1|A_1|1\rangle=1,\quad \langle1|A_2|1\rangle=0,
\]
so
\[
\widetilde K_n^\mathsf{T}A_1\widetilde K_n =\omega_n\omega_n^\mathsf{T}\succeq0,\quad
\widetilde K_n^\mathsf{T}A_2\widetilde K_n =0\succeq0.
\]
Taking $\lambda_1=\lambda_2=0$ for all $n,a$, Theorem~\ref{thm:pi_order_real_operation} implies that $\Phi$ is a $\pi$-ORO.

Each $K_n$ can only have nonzero entries in its second row, so $K_{n,ij}=0$ whenever $i\neq2$.
It follows that
\[
K_{n,\pi^{-1}(r),\pi^{-1}(j)} K_{n,\pi^{-1}(s),\pi^{-1}(j)}=0
\]
for all $n$, and summing over $n$ yields for $\forall r \neq s$,
\[
\sum_{n=1}^3 K_{n,\pi^{-1}(r),\pi^{-1}(j)} K_{n,\pi^{-1}(s),\pi^{-1}(j)} =0,\quad
\forall j\in\{1,2,3\}.
\]
Condition \eqref{eq:condC} is satisfied, so Theorem~\ref{thm:pi-ORO and pi-OIO-1} tells us that $\Phi$ is also a $\pi$-OIO.
The Kraus operators satisfy the characterization of $\pi$-OIOs given in Ref.~\cite[Thm.1]{Qi2026}.
\end{eg}

The channel in Example~\ref{ex:pi-ORO and pi-OIO-1} belongs to both families, i.e., $\pi\text{-ORO}\cap\pi\text{-OIO}\neq\varnothing$. We now treat the second question.

\begin{thm}\label{thm:pi-ORO and pi-OIO-2}
Let $\Phi$ be a $\pi$-OIO, so each Kraus operator admits the form
\[
K_n=\sum_{i=1}^{d} k_{n,i}\, \ket{f_n(i)}\bra{i|},
\]
satisfying the ordering conditions for $\pi$-OIOs from Ref.~\cite[Thm.1]{Qi2026}.
Then $\Phi$ is also a $\pi$-ORO under the following conditions,
1) $k_{n,i}\in\mathbb{R}$; 2) if $i\neq j$ and $k_{n,\pi^{-1}(i)}k_{n,\pi^{-1}(j)}\neq 0$, then $g_n(i)\neq g_n(j)$, where $g_n=\pi\circ f_n\circ \pi^{-1}$.
\end{thm}
\begin{proof}
For any $\rho\in\mathcal{R}_\pi$, $\rho^\mathsf{T}=\rho$. Set $\tau_n=K_n\rho K_n^\mathsf{T}$. Then $\tau_n^\mathsf{T}=\tau_n$, so $\tau_n$ is real symmetric. It remains to verify the $\pi$-monotonicity of its diagonal entries.
Define $\widetilde\rho = P_\pi \rho P_\pi^\mathsf{T}$ with diagonal entries $p_i=\widetilde\rho_{ii}=\rho_{\pi^{-1}(i)\pi^{-1}(i)}$.
By definition of $\mathcal{R}_\pi$, $p_1\ge p_2\ge\cdots\ge p_d\ge 0$ and $\sum_{i=1}^d p_i=1$.
Let $\Delta(\widetilde\rho)$ denote the diagonal part of $\widetilde\rho$:
\[
\Delta(\widetilde\rho)=\sum_{i=1}^{d} p_i\,\ket{i}\bra{i} \in P_\pi \mathcal{F}_\pi P_\pi^\mathsf{T}.
\]

In the $\pi$-ordered basis,
\[
\widetilde K_n = P_\pi K_n P_\pi^\mathsf{T}
=\sum_{i=1}^{d} k_{n,\pi^{-1}(i)}\,\ket{g_n(i)}\bra{i},
~g_n=\pi\circ f_n\circ\pi^{-1}.
\]
Then
\[
\widetilde \tau_n=\widetilde K_n \widetilde\rho \widetilde K_n^\mathsf{T}
=\sum_{i,j=1}^d k_{n,\pi^{-1}(i)}k_{n,\pi^{-1}(j)}\,\ket{g_n(i)}\bra{g_n(j)}\,\widetilde\rho_{ij}.
\]
Its diagonal entry is
\[
(\widetilde \tau_n)_{aa}
=\langle a|\widetilde \tau_n|a \rangle
=\sum_{\substack{i,j\\ g_n(i)=g_n(j)=a}} k_{n,\pi^{-1}(i)}k_{n,\pi^{-1}(j)}\,\widetilde\rho_{ij}.
\]
Condition 2) says that if $g_n(i)=g_n(j)=a$ and both coefficients are nonzero, then necessarily $i=j$.
Consequently,
\[
(\widetilde \tau_n)_{aa}=\sum_{\{i\,\colon\,g_n(i)=a\}} k_{n,\pi^{-1}(i)}^2\, p_i,
\]
which coincides with $(\widetilde \tau_n)_{aa}
=\big(\widetilde K_n\,\Delta(\widetilde\rho)\,\widetilde K_n^\mathsf{T}\big)_{aa}$.
Since $\Delta(\widetilde\rho)\in P_\pi\mathcal{F}_\pi P_\pi^\mathsf{T}$, we have
$\widetilde K_n\Delta(\widetilde\rho)\widetilde K_n^\mathsf{T}\in \mathbb{R}_+ P_\pi\mathcal{F}_\pi P_\pi^\mathsf{T}$,
whose diagonal entries are nonincreasing:
$(\widetilde \tau_n)_{11}\ge (\widetilde \tau_n)_{22}\ge\cdots\ge (\widetilde \tau_n)_{dd}\ge 0$.
Translating back,
\[
(\tau_n)_{\pi^{-1}(1),\pi^{-1}(1)}
\ge (\tau_n)_{\pi^{-1}(2),\pi^{-1}(2)}
\ge\cdots\ge
(\tau_n)_{\pi^{-1}(d),\pi^{-1}(d)}.
\]
Thus $\tau_n\in\mathbb{R}_+\mathcal{R}_\pi$.
We conclude $K_n\mathcal{R}_\pi K_n^\mathsf{T}\subseteq\mathbb{R}_+\mathcal{R}_\pi$ for all $n$, and since each $K_n$ is real, $\Phi$ is a $\pi$-ORO.
\end{proof}

We demonstrate the theorem above with an example.
\begin{eg}
Let $d=3$, $\pi(1,2,3)=(3,1,2)$, and let $\Phi$ have Kraus operators
\[
K_1=\frac1{\sqrt3} \begin{pmatrix} 1&0&0\\ 0&1&0\\ 0&0&1 \end{pmatrix},\quad
K_2=\frac1{\sqrt3} \begin{pmatrix} 1&0&0\\ 0&-1&0\\ 0&0&1 \end{pmatrix}, \]
\[
K_3=\frac1{\sqrt3} \begin{pmatrix} 1&0&0\\ 0&1&0\\ 0&0&-1 \end{pmatrix}.
\]
where all $K_n$ are real with $\sum_{n=1}^3K_n^\mathsf{T}K_n=I_3$.
Each Kraus operator can be written as $K_n=\sum_{i=1}^3 k_{n,i}\ket{f_n(i)}\bra{i}$ with $f_n(i)=i$ for all $i$.
Hence $g_n=\pi\circ f_n\circ\pi^{-1} =\operatorname{id}$, i.e., $g_n(i)=i$.
The coefficient matrix is
\[
(k_{n,i})= \frac1{\sqrt3} \begin{pmatrix} 1&1&1\\ 1&-1&1\\ 1&1&-1 \end{pmatrix}.
\]
Since $f_n(i)=i$, $f_n(i)\neq f_n(j)$ whenever $i\neq j$,
\[
\sum_{\{n:f_n(i)=f_n(j)\}}k_{n,i}^*k_{n,j}=0.
\]
When $i=j$,
\[
\sum_{\{n:f_n(i)=f_n(j)\}}k_{n,i}^*k_{n,j}=\sum_{n=1}^3|k_{n,i}|^2 =3\cdot\frac13=1.
\]
Therefore, $\sum_{\{n:f_n(i)=f_n(j)\}}k_{n,i}^*k_{n,j} =\delta_{ij}$.
Because $g_n=\operatorname{id}$ and $|k_{n,\pi^{-1}(i)}|^2=\frac13$ for all $n$ and $i$, the ordering constraints for $\pi$-OIOs in Ref.~\cite[Thm.1]{Qi2026} are satisfied. Hence, $\Phi$ is a $\pi$-OIO.

The condition 1) in Theorem \ref{thm:pi-ORO and pi-OIO-2} is obvious: $k_{n,i}\in\mathbb R$.
Since $g_n(i)=i$ for $i\neq j$, we have $g_n(i)\neq g_n(j)$.
Thus the implication
\[
k_{n,\pi^{-1}(i)}k_{n,\pi^{-1}(j)}\neq0 \implies g_n(i)\neq g_n(j)
\]
holds. Therefore, $\Phi$ is also a $\pi$-ORO.
One may directly verify that the Kraus operators satisfy the $\pi$-ORO criterion from Theorem~\ref{thm:pi_order_real_operation}.
\end{eg}

\begin{figure}[htbp]
	\centering
	\includegraphics[width=0.45\textwidth]{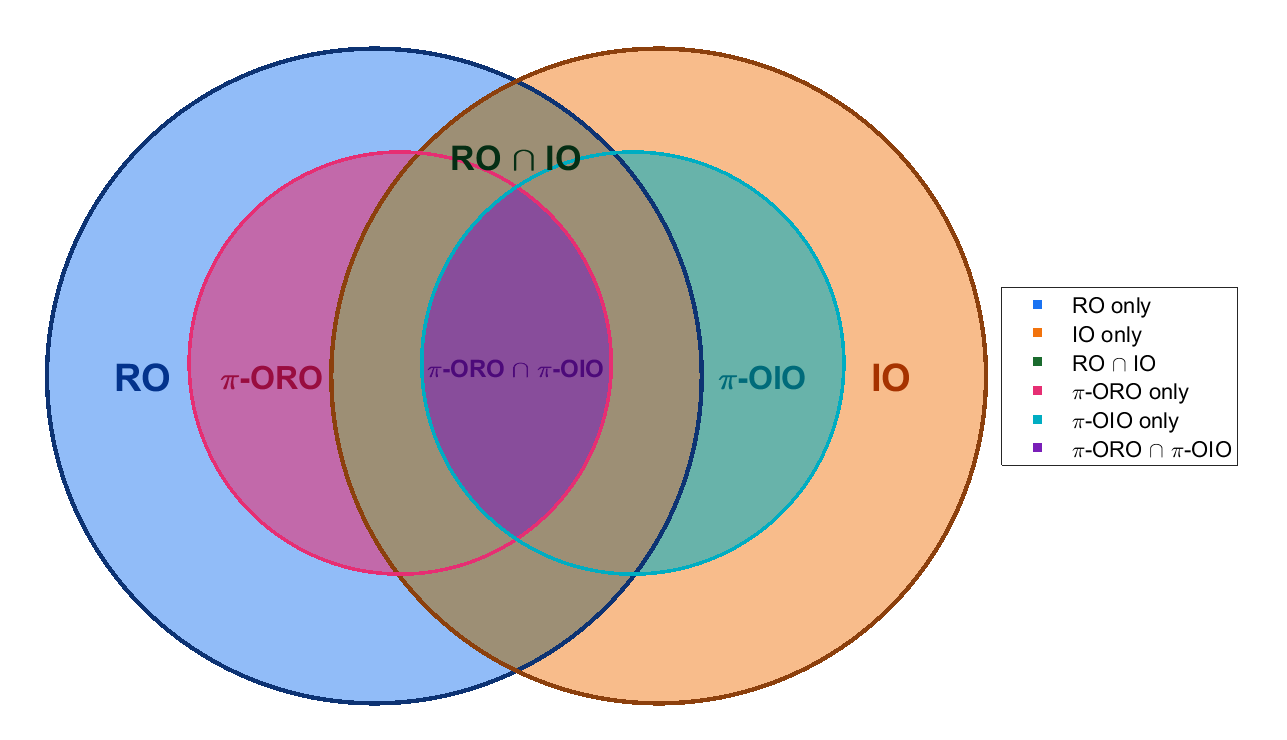}
	\caption{Inclusion relations among $\pi$-OROs, $\pi$-OIOs, real operations (ROs) and incoherent operations (IOs).}
	\label{figure2}
\end{figure}

Theorems~\ref{thm:pi-ORO and pi-OIO-1} and \ref{thm:pi-ORO and pi-OIO-2} together clarify the logical relations among real operations (ROs), incoherent operations (IOs), $\pi$-OROs and $\pi$-OIOs. Theorem~\ref{thm:pi-ORO and pi-OIO-1} gives the necessary and sufficient condition for a given $\pi$-ORO to preserve $\pi$-ordered incoherent states, while Theorem~\ref{thm:pi-ORO and pi-OIO-2} identifies extra conditions under which a given $\pi$-OIO becomes a $\pi$-ORO. The inclusion relations among $\pi$-OROs, $\pi$-OIOs, real operations (ROs) and incoherent operations (IOs) are illustrated in Fig.~\ref{figure2}.

\section{resource measures of  the $\pi$-ordered imaginarity}\label{sect4}
A central problem in the resource theory of \(\pi\)-ordered imaginarity lies in quantitative analysis, aiming to rigorously characterize the resource properties of physical states. In this work, we develop a quantitative theory of \(\pi\)-ordered imaginarity by adopting the approach for imaginary resource measures from Ref.~\cite{Hickey2018,Wu2021}, and further derive a set of efficiently computable quantifiers of \(\pi\)-ordered imaginarity.
	
A nonnegative functional \(\J\) is called a valid \(\pi\)-ordered imaginarity measure if it fulfills the following axioms: (A) Nonnegativity: \(\J(\rho)\ge 0\), and \(\J(\rho)=0\) for all \(\rho\in\cR_\pi\). (B) Monotonicity: For any quantum state \(\rho\in\D(\cH)\) and any \(\pi\)-ordered real operation \(\Phi\) on \(\cB(\cH)\),
\(\J\big(\Phi(\rho)\big)\le \J(\rho).\) (C) Probabilistic monotonicity: \(\sum_n p_n \J(\rho_n)\le \J(\rho)\), where \(p_n=\mathrm{Tr}(K_n \rho K_n^\dagger)\) and \(\rho_n=\frac{1}{p_n}K_n \rho K_n^\dagger\), for all Kraus sets \(\{K_n\}\) such that \(\sum_n K_n^\dagger K_n=I\), \(\langle i |K_n|j\rangle\in\mathbb{R}\) for all \(i,j\), and \(K_n\rho K_n^\dagger\in\cR_\pi\). (D) Convexity: For any ensemble \(\{p_n,\rho_n\}\),
\(\J\Big(\sum_n p_n \rho_n\Big)\le \sum_n p_n \J(\rho_n).\)

Note that axioms (C) and (D) jointly imply axiom (B). In Ref.~\cite{XueQIP383}, the authors considered the following additional condition: (E) Additivity for direct-sum states: \(\J\big(p\rho_1\oplus (1-p)\rho_2\big)= p \J(\rho_1)+(1-p)\J(\rho_2).\) It was proven that the combinations of (B) and (E) are equivalent to the pair of axioms (C) and (D)~\cite{XueQIP383}.

In the following, we propose several natural candidates for \(\pi\)-ordered imaginarity measures.

\subsection{The $\pi$-ordered imaginarity measure induced by relative entropy} 
	
The quantum relative entropy is widely employed in quantum information theory \cite{Chen2025,Baumgratz2014}. Recall that the relative entropy between two quantum states $\rho$ and $\sigma$ is defined as
\[
S(\rho\|\sigma)=\mathrm{Tr}(\rho\log\rho)-\mathrm{Tr}(\rho\log\sigma)
=-S(\rho)-\mathrm{Tr}(\rho\log\sigma),
\]
where $S(\rho)=-\mathrm{Tr}(\rho\log\rho)$ denotes the von Neumann entropy.
For any quantum state $\rho\in\D(\cH)$, we define
\begin{equation}\label{eq:relative entropy}
\J^\pi_S(\rho)=\min_{\sigma\in\cR_\pi} S(\rho\|\sigma)
=\min_{\sigma\in\cR_\pi}\Big\{\mathrm{Tr}(\rho\log\rho)-\mathrm{Tr}(\rho\log\sigma)\Big\},
\end{equation}
where the minimization runs over all $\pi$-ordered real states.
Obviously, $\J^\pi_S(\rho)\ge 0$ for all $\rho\in\D(\cH)$, and $\J^\pi_S(\rho)=0$ if and only if $\rho\in\cR_\pi$.

As established in Ref.~\cite{Baumgratz2014}, the quantum relative entropy $S(\rho\|\sigma)$ satisfies the data-processing inequality:
$S(\rho\|\sigma)\ge S(\Phi(\rho)\|\Phi(\sigma))$ for arbitrary quantum operations $\Phi$.
For any $\pi$-ordered real operation $\Phi$, we then write
\[
\begin{aligned}
\J^\pi_S(\rho) &= S(\rho\|\sigma^*) \ge S\big(\Phi(\rho)\big\|\Phi(\sigma^*)\big) \\
&\ge \min_{\sigma\in\cR_\pi} S\big(\Phi(\rho)\|\sigma\big) = \J^\pi_S\big(\Phi(\rho)\big),
\end{aligned}
\]
where $\sigma^*$ is the $\pi$-ordered real state attaining the minimum.
This shows that $\J^\pi_S$ obeys the axiom (B).

We next use the strong contractivity of relative entropy under quantum operations \cite{lindblad1975entropy}:
$S(\rho\|\sigma)\ge \sum_n q_n S(\rho_n\|\sigma_n)$,
where $\rho_n=\frac{K_n\rho K_n^\dagger}{q_n}$, $\sigma_n=\frac{K_n\sigma K_n^\dagger}{p_n}$,
$q_n=\mathrm{Tr}(K_n\rho K_n^\dagger)$ and $p_n=\mathrm{Tr}(K_n\sigma K_n^\dagger)$.
It follows that
\[
\begin{aligned}
\J^\pi_S(\rho) &= S(\rho\|\sigma^*) \ge \sum_n q_n S(\rho_n\|\sigma_n^*) \\
&\ge \sum_n q_n \min_{\sigma\in\cR_\pi} S(\rho_n\|\sigma)
=\sum_n q_n \J^\pi_S(\rho_n),
\end{aligned}
\]
with $\sigma_n^*=\frac{K_n\sigma^* K_n^\dagger}{p_n}$ and $\sigma_n^*\in\cR_\pi$.
Hence the axiom (C) holds for $\J^\pi_S$.

Finally, $S(\rho\|\sigma)$ is jointly convex \cite{ruskai2002inequalities}:
\[
S\Big(\sum_n q_n \rho_n \Big\|\sum_n q_n \sigma_n\Big)
\le \sum_n q_n S(\rho_n\|\sigma_n).
\]
This property implies the axiom (D),
\[
\begin{aligned}
\sum_n q_n \J^\pi_S(\rho_n)
&=\sum_n q_n \min_{\sigma_n\in\cR_\pi} S(\rho_n\|\sigma_n)
=\sum_n q_n S(\rho_n\|\sigma_n^*) \\
&\ge S\Big(\sum_n q_n \rho_n \Big\|\sum_n q_n \sigma_n^*\Big)
\ge \J^\pi_S\Big(\sum_n q_n \rho_n\Big),
\end{aligned}
\]
where $\sigma_n^*$ denotes the $\pi$-ordered real state that minimizes $S(\rho_n\|\sigma_n)$.

In summary, $\J^\pi_S$ satisfies the axioms (A), (B), (C) and (D), and is therefore a valid $\pi$-ordered imaginarity measure.

Remarkably, for the qubit case ($d=2$), $\J^\pi_S$ admits a closed-form expression.
Let $\rho$ be an arbitrary qubit state,
\[
\rho=\frac12(I+r_x\sigma_x+r_y\sigma_y+r_z\sigma_z)=\frac12\begin{pmatrix}
1+r_z & r_x-i r_y\\
r_x+i r_y & 1-r_z
\end{pmatrix},
\]
with $r_x^2+r_y^2+r_z^2\le1$. Define $\rho_{\mathrm{R}}$ as the real part of $\rho$,
\[
\rho_{\mathrm{R}} =\frac{\rho+\rho^*}{2} = \frac12\begin{pmatrix}
1+r_z & r_x\\
r_x & 1-r_z
\end{pmatrix}.
\]
For $d=2$, any free state $\sigma=\frac12(I+s_x\sigma_x+s_z\sigma_z)\in\cR_\pi$ falls into two cases determined by $\pi$:
$\pi=\mathrm{id}$ enforces $\sigma_{11}\ge \sigma_{22}$, whereas $\pi=(12)$ requires $\sigma_{22}\ge \sigma_{11}$.
Introduce the boundary state
\[
\sigma_\partial=\frac12(I+r_x\sigma_x)=\frac12\begin{pmatrix}
1 & r_x\\
r_x & 1
\end{pmatrix}.
\]
Then
\begin{equation}\label{eq:relative entropy d=2}
\J_S^\pi(\rho)=
\begin{cases}
S(\rho_{\mathrm R})-S(\rho), & \rho_{\mathrm R}\in\cR_\pi,\\[2mm]
S(\sigma_\partial)-S(\rho), & \rho_{\mathrm R}\notin\cR_\pi,
\end{cases}
\end{equation}
which furnishes an explicit formula for evaluating $\J_S^\pi(\rho)$ for any qubit state $\rho$.
A proof of Eq.~(\ref{eq:relative entropy d=2}) is given in Appendix~\ref{app:relative entropy}.

\subsection{The $\pi$-ordered imaginarity measure induced by fidelity}
The fidelity is a central symmetric quantifier for assessing the similarity between two quantum states \cite{Vedral1997,Uhlmann1976,Nielsen2012}. It can be used to verify the performance of quantum error correction and provides a mathematical framework for quantum tasks such as quantum state discrimination and entanglement characterization, rendering it an indispensable tool for the practical implementation of quantum technologies \cite{Beale2018,Simmons2024,Gluza2018}.

Recall that the fidelity between two quantum states $\rho$ and $\sigma$ is defined as
\[
F(\rho,\sigma) = \operatorname{Tr}\sqrt{\rho^{\frac12}\sigma\rho^{\frac12}} = \left\|\rho^{\frac12}\sigma^{\frac12}\right\|_{\mathrm{tr}}.
\]
For any quantum state $\rho\in\D(\cH)$, we construct the following fidelity-based $\pi$-ordered imaginarity measure,
\begin{equation}
\J^\pi_F(\rho) = \min_{\sigma\in\cR_\pi}\bigl\{1-F^2(\rho,\sigma)\bigr\},
\end{equation}
where the minimization runs over all $\pi$-ordered real states.

Since the quantity $1-F^2$ obeys contractivity, strong contractivity, and joint convexity \cite{Xiong2019},
$\J^\pi_F(\rho)$ satisfies axioms (A), (B), (C) and (D), and hence constitutes a valid $\pi$-ordered imaginarity measure.

More generally, suppose a distance ${D}$ satisfies contractivity, strong contractivity and joint convexity.
We may define a distance-based $\pi$-ordered imaginarity measure in a unified way:
\begin{equation}\label{eq:imag-measure}
\J^\pi_D(\rho)=\min_{\sigma\in\cR_\pi}{D}(\rho,\sigma),
\end{equation}
where the minimum is taken over all $\pi$-ordered real states in $\cR_\pi$.
Evidently, both the relative-entropy quantifier and the fidelity quantifier introduced above fall into this family of distance-based $\pi$-ordered imaginarity measures.
     
\subsection{Robustness of $\pi$-ordered imaginarity}
	
The robustness quantifiers in resource theory quantify how resilient a quantum state is against perturbations such as decoherence and noise \cite{Jang2006,Ding2024,Monti2018}. These quantifiers can guide the optimization of quantum computing protocols and improve the noise resilience of quantum gates as well as quantum error-correcting codes \cite{Ribeiro2019,Hou2024,Bolt2016}.

Following the definition of the robustness of imaginarity introduced in Ref.~\cite{Wu2021}, we now define the robustness of $\pi$-ordered imaginarity of any quantum state $\rho\in\D(\cH)$ to be
\begin{equation}\label{eq:robust-pi-order}
\J^\pi_R(\rho)=\min_{\tau\in\D(\cH)}\left\{s\ge 0\,\bigg|\, \frac{\rho+s\tau}{1+s}=:\sigma\in\cR_\pi\right\}.
\end{equation}
By construction, $\J^\pi_R(\rho)$ quantifies the minimal amount of mixing required to render $\rho$ a $\pi$-ordered real state. We next verify that $\J^\pi_R(\rho)$ satisfies axioms (A), (B), (C) and (D), so that it is a valid resource quantifier.

From Eq.~(\ref{eq:robust-pi-order}), it is immediate that $\J^\pi_R(\rho)\ge 0$ for all $\rho\in\D(\cH)$, and $\J^\pi_R(\rho)=0$ if and only if $\rho\in\cR_\pi$.

Now consider an arbitrary $\pi$-ordered real operation (\(pi\)-ORO) $\Phi(\cdot)=\sum_n K_n(\cdot)K_n^\dagger$. For a fixed $\rho\in\D(\cH)$, let $\tau^*\in\D(\cH)$ and $\sigma^*\in\cR_\pi$ denote the states attaining the minimum in Eq.~(\ref{eq:robust-pi-order}). Then
\[
\rho=\big[1+\J^\pi_R(\rho)\big]\sigma^*-\J^\pi_R(\rho)\tau^*.
\]
Substituting into the Kraus map yields
\[
\rho_{n}=\frac{K_{n}\rho K_{n}^{\dagger}}{\mathrm{Tr}(K_{n}\rho K_{n}^{\dagger})}=\frac{K_{n}\sigma^{*}K_{n}^{\dagger}+\J^\pi_R(\rho)
K_{n}(\sigma^{*}-\tau^*)K_{n}^{\dagger}}{\mathrm{Tr}(K_{n}\rho K_{n}^{\dagger})}.
\]
Define
\[
s_{n}=\J^\pi_R(\rho)\frac{\mathrm{Tr}(K_{n}\tau^{*}K_{n}^{\dagger})}{\mathrm{Tr}(K_{n}\rho K_{n}^{\dagger})},
\sigma_{n}=\frac{\big[1+\J^\pi_R(\rho)\big]K_{n}\sigma^{*}K_{n}^{\dagger}}
{(1+s_{n})\mathrm{Tr}(K_{n}\rho K_{n}^{\dagger})},\]and 
\[
\tau_{n}=\frac{\J^\pi_R(\rho)K_n\tau^*K_n^\dagger}{s_{n}\mathrm{Tr}(K_{n}\rho K_{n}^{\dagger})}.
\]
We have
\begin{equation}\label{eq:pi-robust}
\rho_{n}=(1+s_{n})\sigma_{n}-s_{n}\tau_{n}.
\end{equation}
We adopt the convention $\rho_{n}=\frac{1}{d}I$ whenever $\mathrm{Tr}(K_{n}\rho K_{n}^{\dagger})=0$. Note that $K_{n}\sigma^{*}K_{n}^{\dagger}\in\cR_\pi$, and $\mathrm{Tr}(\sigma_{n})=\mathrm{Tr}(\tau_{n})=1$. Consequently, $\sigma_{n}\in\cR_\pi$ and $\tau_{n}\in\D(\cH)$. Equation~\eqref{eq:pi-robust} then implies $\J^\pi_R(\rho_{n})\le s_{n}$, i.e.,
$\mathrm{Tr}(K_{n}\rho K_{n}^{\dagger})\J^\pi_R(\rho_{n})\le \J^\pi_R(\rho)\mathrm{Tr}(K_{n}\tau^{*}K_{n}^{\dagger})$.
Summing over $n$, we obtain
\[\begin{array}{llllll}
\sum_{n}q_n\J^\pi_R(\rho_{n})
&=\sum_{n}\mathrm{Tr}(K_{n}\rho K_{n}^{\dagger})\J^\pi_R(\rho_{n})\\
&\le \sum_{n}\J^\pi_R(\rho)\mathrm{Tr}(K_{n}\tau^{*}K_{n}^{\dagger})
=\J^\pi_R(\rho),
\end{array}
\]
which establishes the axiom (C).

We turn next to the convexity. Take any $\rho_{1},\rho_{2}\in\D(\cH)$. Again from Eq.~(\ref{eq:robust-pi-order}), there exist $\tau_{1}^{*},\tau_{2}^{*}\in\D(\cH)$ and $\sigma_{1}^{*},\sigma_{2}^{*}\in\cR_\pi$ such that
$\rho_{1}=\big[1+\J^\pi_R(\rho_{1})\big]\sigma_{1}^{*}-\J^\pi_R(\rho_{1})\tau_{1}^{*}$
and $\rho_{2}=\big[1+\J^\pi_R(\rho_{2})\big]\sigma_{2}^{*}-\J^\pi_R(\rho_{2})\tau_{2}^{*}$.
For any $t\in[0,1]$, set
\[
\sigma=\frac{1}{1+s}\Big\{t\big[1+\J^\pi_R(\rho_{1})\big]\sigma_{1}^{*}+(1-t)\big[1+\J^\pi_R(\rho_{2})\big]\sigma_{2}^{*}\Big\},
\]
\[
\tau=\frac{1}{s}\Big[t\J^\pi_R(\rho_{1})\tau_{1}^{*}+(1-t)\J^\pi_R(\rho_{2})\tau_{2}^{*}\Big],
\]
\[
s=t\J^\pi_R(\rho_{1})+(1-t)\J^\pi_R(\rho_{2}).
\]
One verifies that $\rho=t\rho_{1}+(1-t)\rho_{2}=(1+s)\sigma-s\tau$. Since $\mathrm{Tr}(\sigma)=\mathrm{Tr}(\tau)=1$, we have $\sigma\in\cR_\pi$ and $\tau\in\D(\cH)$. It follows that $\J^\pi_R(\rho)\le s$, or equivalently,
\[
\J^\pi_R\big(t\rho_{1}+(1-t)\rho_{2}\big)\le t\J^\pi_R(\rho_{1})+(1-t)\J^\pi_R(\rho_{2}).
\]
Hence, the axiom (D) holds.

As axioms (C) and (D) jointly imply the axiom (B), we conclude that $\J^\pi_R$ is a valid $\pi$-ordered imaginarity measure.

Direct evaluation of $\J^\pi_R$ via the original definition~\eqref{eq:robust-pi-order} is generally computationally difficult. The following proposition provides a semidefinite programming formulation for computing $\J^\pi_R$, see proof  in Appendix~\ref{app:robustness}.

\begin{prop}\label{prop:pi-robust-2}
For any quantum state $\rho\in\D(\cH)$, the robustness of $\pi$-ordered imaginarity admits the primal-dual semidefinite representation
\begin{align*}
\J^\pi_R(\rho) &= \min_{\Gamma}\big\{\operatorname{Tr}\Gamma -1: \Gamma-\rho \succeq 0,\ \Gamma\in\operatorname{cone}(\cR_\pi)\big\},\\
&= \max_{\substack{Y\succeq 0}}\big\{\operatorname{Tr}(\rho Y)-1: \operatorname{Tr}(\sigma Y)\le 1,\ \forall\sigma\in\cR_\pi\big\},
\end{align*}
where $\operatorname{cone}(\cR_\pi) = \{t\sigma:t\ge0,\ \sigma\in\cR_\pi\}$.
\end{prop}

\begin{eg}
Consider the two-qubit pure state $\rho_\lambda=|\phi_\lambda\rangle\langle\phi_\lambda|$ with $|\phi_\lambda\rangle=\sqrt{\lambda}|00\rangle+ i \sqrt{1-\lambda}|11\rangle$ and $\lambda\in[0,1]$. In the orthogonal basis $\{|00\rangle,|01\rangle,|10\rangle,|11\rangle\}$, $\rho_\lambda$ admits the matrix representation
\[
\begin{bmatrix}
\lambda & 0 & 0 & -i\sqrt{\lambda(1-\lambda)} \\
0 & 0 & 0 & 0 \\
0 & 0 & 0 & 0 \\
i\sqrt{\lambda(1-\lambda)} & 0 & 0 & 1-\lambda
\end{bmatrix},
\]
whose eigenvalues are $\lambda_1=1$ and $\lambda_2=\lambda_3=\lambda_4=0$.

Take the permutation $\pi=(3,1,2,4)$. Direct calculation yields
\[
\J_S^\pi(\rho_\lambda)=
\begin{cases}
\log 4, & 0\le \lambda \le \dfrac{3}{4},\\[4pt]
h_2(\lambda) + \lambda\log 3, & \dfrac{3}{4}< \lambda \le 1,
\end{cases}
\]
where $h_2(\lambda) = -\lambda\log\lambda -(1-\lambda)\log(1-\lambda)$, and
\[
\J_R^\pi(\rho_\lambda)=
\begin{cases}
3, & 0\le \lambda \le 3/4,\\[4pt]
2\lambda + 2\sqrt{3\lambda(1-\lambda)}, & 3/4< \lambda \le 1.
\end{cases}
\]

Recall that the standard relative-entropy imaginarity quantifier $\cI_{\text{rel}}$ \cite{XueQIP383} and the standard robustness-of-imaginarity quantifier $\cI_R$ \cite{Wu2021} are defined respectively as
\[
\cI_{\text{rel}}(\rho) = \min_{\sigma\in\cR} S(\rho\|\sigma)= S\big(\operatorname{Re}(\rho)\big) - S(\rho),
\]
and
\[
\cI_R(\rho) = \min_{\tau\in\D(\cH)}
\left\{ s\ge0 \,\bigg|\, \frac{\rho+s\tau}{1+s}=:\sigma\in\cR \right\}= \frac12 \lVert \rho - \rho^\mathsf{T} \rVert_1,\quad \forall \rho\in\D(\cH),
\]
where $\cR$ denotes the set of all real quantum states in $\D(\cH)$ and the trace norm is given by $\lVert M \rVert_1 = \operatorname{Tr}\sqrt{M^\dagger M}$.

Figure~\ref{figure3} displays $\J_S^\pi(\rho_\lambda)$ (dashed red), $\J_R^\pi(\rho_\lambda)$ (solid blue), $\cI_{\mathrm{rel}}(\rho_\lambda)$ (dotted cyan), and $\cI_R(\rho_\lambda)$ (purple triangles) as functions of $\lambda$ for the pure state $\rho_\lambda$ under permutation $\pi=(3,1,2,4)$. One observes that both $\J_S^\pi(\rho_\lambda)$ and $\J_R^\pi(\rho_\lambda)$ remain constant initially and then decrease as $\lambda$ increases. Moreover, $\J_S^\pi(\rho_\lambda)>0$ and $\J_R^\pi(\rho_\lambda)>0$ for all $\lambda\in[0,1]$, which implies that $\rho_\lambda$ is never a $\pi$-ordered real (free) state. In contrast, $\cI_R(\rho_\lambda)$ in the left panel and $\cI_{\mathrm{rel}}(\rho_\lambda)$ in the right panel first rise and then decay. At the endpoints $\lambda=0$ and $\lambda=1$, $\rho_\lambda$ reduces to a real state. It is further evident that $\J_S^\pi(\rho_\lambda)$ and $\J_R^\pi(\rho_\lambda)$ are uniformly larger than their conventional imaginarity counterparts. This observation demonstrates that $\rho_\lambda$ carries a strictly larger amount of $\pi$-ordered imaginarity than ordinary imaginarity. Within this permutation-ordering framework, the system supports a richer class of quantum resources than those captured by standard imaginarity. This fact underscores that the resource theory of $\pi$-ordered imaginarity is independent of the conventional imaginarity framework, and lays the theoretical groundwork for exploring and harnessing new families of quantum resources.
\begin{figure}[htbp]
\centering
\includegraphics[width=0.5\textwidth]{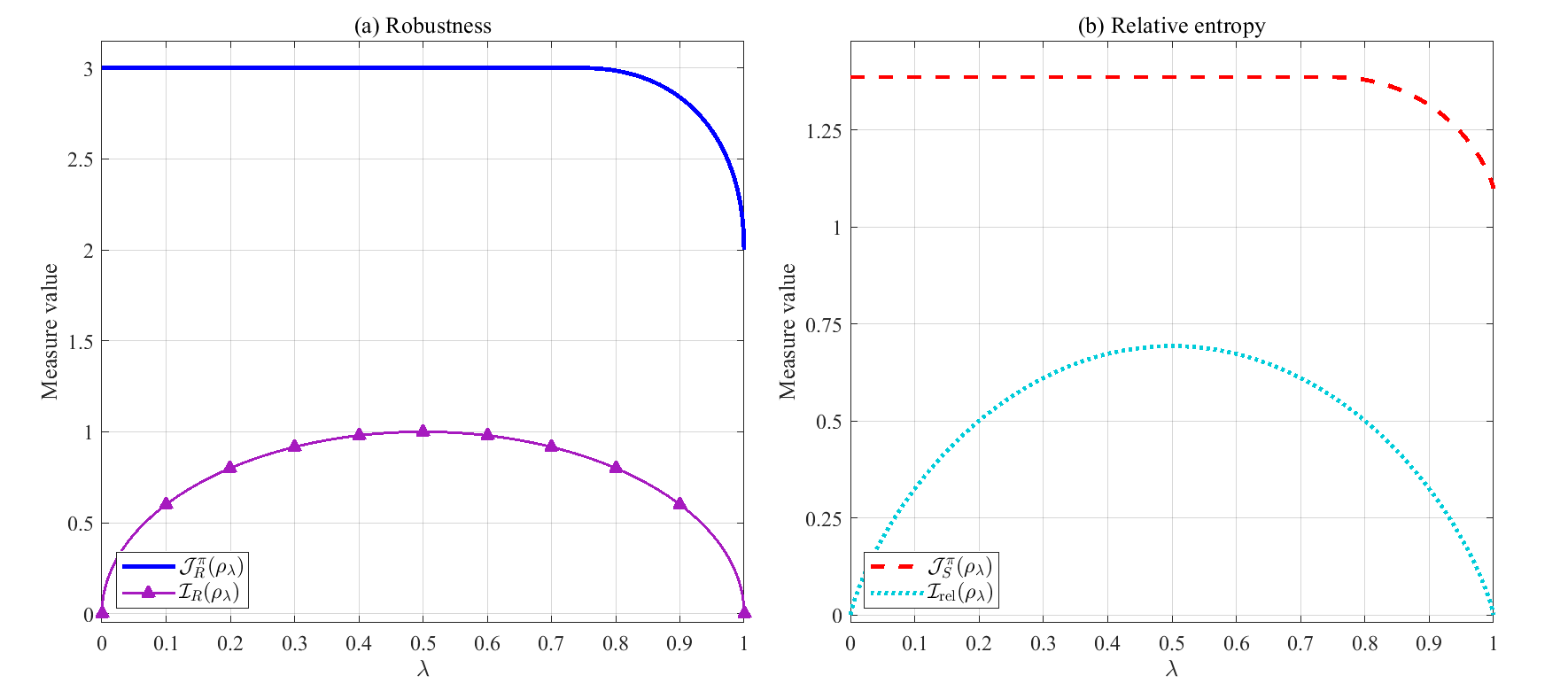}
\caption{Parameter-measure plot of the two-qubit pure state under $\boldsymbol{\pi}=(3,1,2,4)$. In the left subplot, the solid blue curve corresponds to the robustness quantifier $\mathcal{J}_R^\pi(\rho_\lambda)$, and the purple curve with triangular markers is the conventional robustness of imaginarity $\cI_R(\rho_\lambda)$. In the right subplot, the dashed red curve represents the relative-entropy quantifier $\J_S^\pi(\rho_\lambda)$, while the dotted cyan curve shows the conventional relative-entropy imaginarity $\cI_{\mathrm{rel}}(\rho_\lambda)$.}
\label{figure3}
\end{figure}
\end{eg}

The distance-based $\pi$-ordered coherence quantifier $\cC_\D^\pi(\rho)= \min_{\delta\in\mathcal F_\pi}\D(\rho,\delta)$ was introduced in Ref.~\cite{Qi2026}. Since $\F_\pi\subseteq\cR_\pi$, it immediately follows that $\J_\D^\pi(\rho) \le \cC_\D^\pi(\rho)$. This shows that the $\pi$-ordered coherence of a state is at least as large as its $\pi$-ordered imaginarity.

\begin{dfn}
Let $\D$ be any metric satisfying the following three properties:
(i) positivity: $\D(\rho, \sigma) > 0$ for $\rho \neq \sigma$ and $\D(\rho, \rho) = 0$;
(ii) symmetry: $\D(\rho, \sigma) = \D(\sigma, \rho)$;
(iii) triangle inequality: $\D(\rho, \sigma) + \D(\sigma, \tau) \ge \D(\rho, \tau)$.
We define the $\pi$-ordered residual coherence as
\[
\Gamma_\D^\pi(\rho)= \min_{\substack{\sigma\in\operatorname{Argmin}_{\mathcal R_\pi}\D(\rho,\sigma)\\ \delta\in\operatorname{Argmin}_{\mathcal F_\pi}\D(\rho,\delta)}} \D(\sigma,\delta),
\]
where $\operatorname{Argmin}$ denotes the set of minimizers achieving the minimal distance. Clearly, $\Gamma_\D^\pi(\rho)\ge 0$, with equality if and only if $\sigma=\delta$.
\end{dfn}

\begin{figure}[htbp]
\centering
\includegraphics[width=0.4\textwidth]{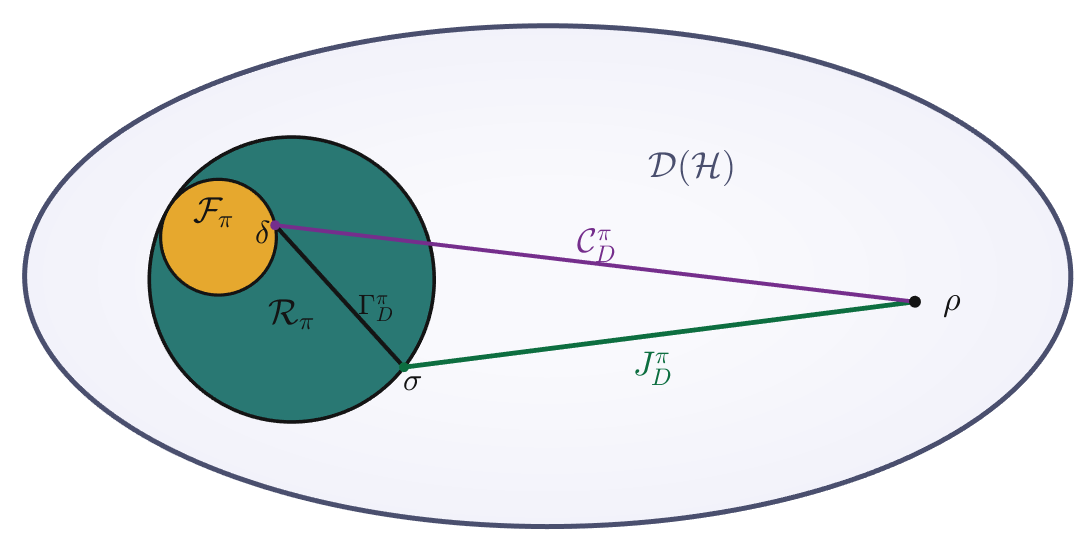}
\caption{Geometric illustration for $\pi$-ordered resource quantifiers. The total $\pi$-ordered coherence of a quantum state is bounded above by the sum of its $\pi$-ordered imaginarity (distance to the nearest $\pi$-ordered real state) and its $\pi$-ordered residual coherence (distance between the nearest $\pi$-ordered real state and the nearest $\pi$-ordered incoherent state). This demonstrates that the $\pi$-ordered imaginarity forms a resource component contained within $\pi$-ordered coherence.}
\label{figure4}
\end{figure}

Choose minimizers $\sigma_\rho^*$ and $\delta_\rho^*$ such that
$\cC_\D^\pi(\rho) = \D(\rho,\delta_\rho^*)$, $\J_\D^\pi(\rho) = \D(\rho,\sigma_\rho^*)$, and $\Gamma_\D^\pi(\rho)=\D(\sigma_\rho^*,\delta_\rho^*)$.
Applying the triangle inequality,
$\cC_\D^\pi(\rho) =\D(\rho,\delta_\rho^*)
\le \D(\rho,\sigma_\rho^*) + \D(\sigma_\rho^*,\delta_\rho^*)
= \J_\D^\pi(\rho) + \Gamma_\D^\pi(\rho)$,  we obtain the chain
$\J_\D^\pi(\rho) \le \cC_\D^\pi(\rho) \le \J_\D^\pi(\rho) + \Gamma_\D^\pi(\rho)$,
which encodes a fundamental tradeoff among the three quantifiers $\J_\D^\pi$, $\cC_\D^\pi$ and $\Gamma_\D^\pi$, see Fig.~\ref{figure4}.
Combined with the above statement, we obtain the following inequality.
\begin{thm}
For any quantum state $\rho$ and any metric $\D$ satisfying conditions (i)--(iii),
\begin{equation}\label{eq:11}
\cC_\D^\pi(\rho) \le \J_\D^\pi(\rho) + \Gamma_\D^\pi(\rho).
\end{equation}
\end{thm}

We illustrate the inequality~\eqref{eq:11} by using the trace distance $\mathcal D(\rho,\sigma)=\frac12\|\rho-\sigma\|_1$, which satisfies properties (i)–(iii). Set $\pi=\mathrm{id}$ and consider the qubit state
\[
\rho_t=\frac12 \begin{pmatrix} 1&t(1-i)\\ t(1+i)&1 \end{pmatrix},\qquad 0<t\leq\frac1{\sqrt2}.
\]
Any state $\delta\in\mathcal F_\pi$ can be parametrized as
\[
\delta_u=\frac12 \begin{pmatrix} 1+u&0\\ 0&1-u \end{pmatrix},\qquad 0\leq u\leq1.
\]
Direct computation gives $\mathcal D(\rho_t,\delta_u)
=\frac12\sqrt{2t^2+u^2}$. The minimizer is therefore $\delta^*=\tfrac12 I$, yielding $\mathcal C_{\mathcal D}^{\pi}(\rho_t)=\tfrac{t}{\sqrt2}$.

Next, any $\sigma\in\mathcal R_\pi$ takes the form
\[
\sigma_{v,u}=\frac12 \begin{pmatrix} 1+u&v\\ v&1-u \end{pmatrix},
\]
with $u\ge0$ and $v^2+u^2\le1$. One finds
\[
\mathcal D(\rho_t,\sigma_{v,u}) =\frac12\sqrt{(t-v)^2+t^2+u^2}.
\]
Hence, the optimal real state is
\[
\sigma^*=\frac12 \begin{pmatrix} 1&t\\ t&1 \end{pmatrix},
\]
and $\mathcal J_{\mathcal D}^{\pi}(\rho_t)=\tfrac t2$.
Moreover,
\[
\Gamma_{\mathcal D}^{\pi}(\rho_t)=\mathcal D(\sigma^*,\delta^*)=\frac12\left\| \begin{pmatrix} 0&t/2\\ t/2&0 \end{pmatrix} \right\|_1 =\frac t2.
\]
Substituting these values confirms
\[
\mathcal C_{\mathcal D}^{\pi}(\rho_t) =\frac t{\sqrt2} < t =\mathcal J_{\mathcal D}^{\pi}(\rho_t) +\Gamma_{\mathcal D}^{\pi}(\rho_t).
\]

The inequality (\ref{eq:11}) reveals a fundamental geometric constraint linking $\pi$-ordered coherence and $\pi$-ordered imaginarity: the total $\pi$-ordered coherence is bounded from above by the sum of $\pi$-ordered imaginarity and $\pi$-ordered residual coherence. Equality in~\eqref{eq:11} holds if and only if there exists a minimizing pair $(\sigma_\rho^*,\delta_\rho^*)$ satisfying $\mathcal{D}(\rho,\delta_\rho^*)=\mathcal{D}(\rho,\sigma_\rho^*)+\mathcal{D}(\sigma_\rho^*,\delta_\rho^*)$.
In particular, every $\pi$-ordered real state saturates this bound: for such $\rho$, $\sigma_\rho^*=\rho$ and $\mathcal{J}_{\mathcal{D}}^\pi(\rho)=0$, which forces $\Gamma_{\mathcal{D}}^\pi(\rho)=\mathcal{C}_{\mathcal{D}}^\pi(\rho)$.
Equality is also achieved when the two minimizers coincide, i.e., $\sigma_\rho^*=\delta_\rho^*$. In this case, $\Gamma_{\mathcal{D}}^\pi(\rho)=0$ and $\mathcal{C}_{\mathcal{D}}^\pi(\rho)=\mathcal{J}_{\mathcal{D}}^\pi(\rho)$.
We emphasize that this upper bound arises purely from geometric distance relations and, in general, cannot be interpreted as an exact additive decomposition into two independent resource contributions.

The robustness quantifiers, however, are not standard metrics. Hence, an analogous simple inequality does not follow automatically. Recall that the robustness of $\pi$-ordered coherence $\cC_R^\pi$, defined in Ref.~\cite{Qi2026}, reads
\[
\mathcal{C}_R^\pi(\rho) = \min_{\eta\in\mathcal{D}(\mathcal{H})} \left\{ s\ge 0 \,\bigg|\, \frac{\rho+s\eta}{1+s}=:\delta \in \mathcal{F}_\pi \right\}.
\]
Following the definition~\eqref{eq:robust-pi-order}, let $j=\J_R^\pi(\rho)$ and the optimal decomposition be $\rho = (1+j)\sigma^* - j\tau^*$, where $\sigma^*\in\mathcal{R}_\pi$ and $\tau^*\in\D(\mathcal{H})$.
We introduce the robustness-based residual coherence,
\[
\gamma_R^\pi(\rho)=\inf_{\sigma^*\in\operatorname{Opt}_{\mathcal{R}_\pi}(\rho)} \mathcal{C}_R^\pi(\sigma^*),
\]
which is the minimal $\pi$-ordered coherence robustness required to map an optimal $\pi$-ordered real state of $\rho$ into a $\pi$-ordered incoherent state.
Here $\operatorname{Opt}_{\mathcal{R}_\pi}(\rho)$ denotes the set of optimal $\pi$-ordered real states for $\rho$.

Take the optimal decomposition for $\sigma^*$, $\sigma^* = (1+\gamma_R^\pi)\delta^* - \gamma_R^\pi\eta^*$,
with $\delta^*\in\mathcal{F}_\pi$ and $\eta^*\in\mathcal{D}(\mathcal{H})$. Substituting it into the expression for $\rho$,
we have
\begin{align*}
\rho &= (1+j)\big[(1+\gamma_R^\pi)\delta^*-\gamma_R^\pi\eta^*\big]-j\tau^* \\
&= (1+j)(1+\gamma_R^\pi)\delta^* - \big[(1+j)\gamma_R^\pi\eta^* + j\tau^*\big].
\end{align*}
Set $A=(1+j)(1+\gamma_R^\pi)$ and $B=(1+j)\gamma_R^\pi\eta^* + j\tau^*$, so that $\rho=A\delta^*-B$.
Since
\[
\operatorname{Tr}(B)=(1+j)\gamma_R^\pi+j=(1+j)(\gamma_R^\pi+1)-1=A-1,
\]
we define $\xi=\tfrac{B}{\operatorname{Tr}(B)}$, which satisfies $B=(A-1)\xi$. Consequently,
\[
\rho=A\delta^*-(A-1)\xi=\big(1+(A-1)\big)\delta^*-(A-1)\xi.
\]
This is a valid robustness decomposition that converts $\rho$ into a $\pi$-ordered incoherent state with parameter $s=A-1$. It follows that
\begin{align*}
\mathcal{C}_R^\pi(\rho) &\le A-1 = (1+\mathcal{J}_R^\pi(\rho))(1+\gamma_R^\pi(\rho))-1 \\
&=\mathcal{J}_R^\pi(\rho)+(1+\mathcal{J}_R^\pi(\rho))\gamma_R^\pi(\rho).
\end{align*}
Summarizing the above, we arrive at the following inequality.
\begin{thm}
For any quantum state $\rho$,
\begin{equation}\label{eq:12}
\mathcal{C}_R^\pi(\rho)\le\mathcal{J}_R^\pi(\rho)+(1+\mathcal{J}_R^\pi(\rho))\gamma_R^\pi(\rho).
\end{equation}
\end{thm}

We validate inequality~\eqref{eq:12} on an explicit example. Let $\pi=\mathrm{id}$ and take the qubit state
\[
\rho=\frac13 \begin{pmatrix} 2&1-i\\ 1+i&1 \end{pmatrix}.
\]
From the definitions of $\mathcal{C}_R^\pi(\rho)$, $\mathcal{J}_R^\pi(\rho)$ and $\gamma_R^\pi(\rho)$, direct evaluation yields
\[
\mathcal C_R^\pi(\rho)=\frac{2\sqrt2}{3},\qquad \mathcal J_R^\pi(\rho)=\frac23,\qquad \gamma_R^\pi(\rho)=\frac25.
\]
One then verifies that
\[
\mathcal C_R^\pi(\rho)=\frac{2\sqrt{2}}{3} < \frac43 = \mathcal J_R^\pi(\rho) +\bigl(1+\mathcal J_R^\pi(\rho)\bigr)\gamma_R^\pi(\rho).
\]

\section{Conclusion}
\label{sec5}

We have developed the resource theory of \(\pi\)-ordered imaginarity, a refined framework that simultaneously treats two distinct quantum resources: imaginarity associated with complex phases of density matrices and resourcefulness originating from the population misordering. Unlike the conventional imaginarity theory, which merely identifies real density matrices as free states, our framework imposes an additional permutation constraint \(\pi\) on the diagonal populations of free states. This construction is motivated by the central role of population ordering and incoherent ergotropy in quantum battery thermodynamics, where population redistribution and population inversion contribute directly to the extractable work.

We have analyzed the mathematical structure of \(\pi\)-ordered real free states using majorization theory and the Schur–Horn theorem. Within the Schur–Horn polytope containing all admissible diagonal vectors compatible with a fixed eigenvalue spectrum, the \(\pi\)-order condition singles out a structured subset of real states. We have also established the geometric correspondence between \(\pi\)-ordered real states and \(\pi\)-ordered incoherent states: every \(\pi\)-ordered real state can be unitarily rotated into a diagonal state belonging to \(\mathcal F_\pi\) while preserving its full eigenvalue spectrum. We have characterized the admissible \(\pi\)-ordered real operations and examined their overlap with \(\pi\)-ordered incoherent operations, clarifying how the reality and population-order constraints jointly restrict allowed state transformations.

We have formulated the general definition of \(\pi\)-ordered imaginarity monotones and constructed concrete quantifiers based on relative entropy, fidelity and robustness, explored the geometric relation between \(\pi\)-ordered imaginarity and \(\pi\)-ordered coherence, and derived inequalities connecting their respective robustness measures. These results demonstrate how population ordering reshapes the resource landscape of imaginarity and distinguishes states that are equivalent under the standard imaginarity criterion.

Our framework extends the imaginarity resource theory and yields a flexible toolbox parameterized by an arbitrary permutation \(\pi\). Several promising directions remain open for future research. It would be worthwhile to investigate deterministic and probabilistic state conversion and distillation protocols under \(\pi\)-ordered real operations, as well as the operational advantages of \(\pi\)-ordered imaginarity in work-extraction and quantum battery charging protocols. Extending the present state-based formalism to the case of quantum channels, in parallel with the existing channel imaginarity theories, constitutes another natural direction. Moreover, further study on interplay among \(\pi\)-ordered imaginarity, \(\pi\)-ordered coherence and other ordered quantum resources may bring new insights into quantum metrology and quantum thermodynamics.

\section*{Acknowledgments:}
Shao-Ming Fei acknowledges the financial support from specific research fund of the
Innovation Platform for Academicians of Hainan Province.
\section*{Data availability}
No data were created or analyzed in this study.

\appendix

\section{Proof of Theorem~\ref{thm:pi_order_real_operation}}\label{app:proof of theorem 2}
Take an arbitrary $\rho\in\mathcal{R}_\pi$. Since each $K_n$ is real, for any real vector $z$ we have
$z^\mathsf{T}K_n\rho K_n^\mathsf{T}z
=(K_n^\mathsf{T}z)^\mathsf{T}\rho(K_n^\mathsf{T}z)\ge0$.
Hence, $K_n\rho K_n^\mathsf{T}\succeq0$.
The inclusion $K_n \mathcal{R}_\pi K_n^\mathsf{T} \subseteq \mathcal{R}_\pi$ is therefore equivalent to $K_n \mathcal{R}_\pi K_n^\mathsf{T} \subseteq \mathbb{R}_+\mathcal{R}_\pi$, where $\mathbb{R}_+\mathcal{R}_\pi$ denotes the associated free cone.

Let $q_a = \pi^{-1}(a)$ for $a=1,2,\dots,d$, and define
$F_a^{(\pi)} = \ket{q_a}\bra{q_a} - \ket{q_{a+1}}\bra{q_{a+1}}$.
The free cone is
\[
\mathcal{C}_\pi := \mathbb{R}_+\mathcal{R}_\pi= \bigl\{ \rho\in\operatorname{Sym}(\mathcal{H}):\; \rho\succeq 0,\; \mathrm{Tr}\big(F_a^{(\pi)} \rho\big)\ge 0,\bigr\}.
\] where $a=1,\dots,d-1$ and
the constraint $\mathrm{Tr}(F_a^{(\pi)}\rho)\ge0$ is exactly the diagonal ordering condition $\rho_{q_a q_a}-\rho_{q_{a+1}q_{a+1}}\ge0$.

Because the map $X\mapsto K_n X K_n^\mathsf{T}$ is affine linear, we obtain the following equivalence,
\[
K_n \mathcal{R}_\pi K_n^\mathsf{T} \subseteq \mathbb{R}_+\mathcal{R}_\pi,\quad \forall n
\;\Longleftrightarrow\;
K_n \mathcal{C}_\pi K_n^\mathsf{T} \subseteq \mathcal{C}_\pi,\quad \forall n.
\]
This means that for every $\rho\in\mathcal{C}_\pi$ and every Kraus operator $K_n$, the output $K_n\rho K_n^\mathsf{T}$ must belong to $\mathcal{C}_\pi$. As the real symmetry and positive semidefiniteness of $K_n\rho K_n^\mathsf{T}$ are already guaranteed, it suffices to enforce
$\mathrm{Tr}\big(F_a^{(\pi)} K_n \rho K_n^\mathsf{T}\big)
= \mathrm{Tr}\big(K_n^\mathsf{T} F_a^{(\pi)} K_n \rho\big) \ge 0$.
We thus arrive at the equivalent requirement,
\[
\forall \rho\in\mathcal{C}_\pi,\; \mathrm{Tr}\big(K_n^\mathsf{T} F_a^{(\pi)} K_n \rho\big)\ge 0,\quad \forall n,a.
\]

\begin{figure}[htbp]
    \centering
    \includegraphics[width=0.4\textwidth]{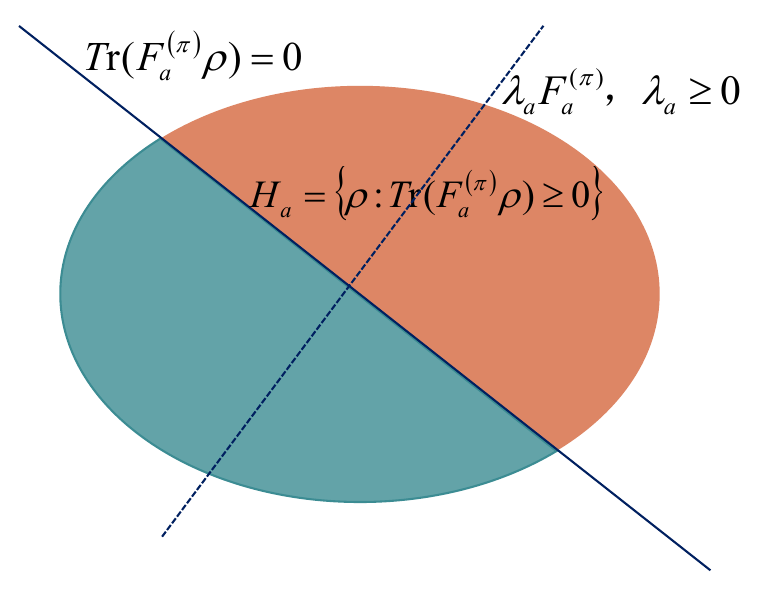}
    \caption{Geometric illustration of the half-space cone $H_a$.}
    \label{figure1}
\end{figure}

Recall the definition of the dual cone \cite{Boyd2004}:
\[
\mathcal{C}_\pi^* = \bigl\{ \sigma\in\operatorname{Sym}(\mathcal{H}): \mathrm{Tr}(\sigma \rho)\ge0,\; \forall \rho\in\mathcal{C}_\pi \bigr\}.
\]
By definition, $\sigma\in\mathcal{C}_\pi^*$ if its Hilbert–Schmidt inner product with every element of $\mathcal{C}_\pi$ is nonnegative.
The preceding condition is equivalent to
$K_n^\mathsf{T} F_a^{(\pi)} K_n \in \mathcal{C}_\pi^*$, $\forall n,a$.
Write
\[
\mathcal{C}_\pi = S_+^d \bigcap_{a=1}^{d-1} H_a,
\]
where $S_+^d=\{\rho\in\operatorname{Sym}(\mathcal{H}): \rho\succeq0\}$ is the positive-semidefinite cone, and
$H_a = \bigl\{ \rho: \mathrm{Tr}(F_a^{(\pi)} \rho)\ge0 \bigr\}$.
Each $H_a$ is a half-space cone separated from its complement by the hyperplane $\mathrm{Tr}(F_a^{(\pi)}\rho)=0$, with normal vector $F_a^{(\pi)}$, see Fig.~\ref{figure1}.

The cone $S_+^d$ is self-dual, i.e., $(S_+^d)^*=S_+^d$ \cite{Boyd2004}.
The dual of a half-space cone is the ray spanned by its normal vector: $H_a^* = \{\lambda_a F_a^{(\pi)}: \lambda_a\ge0\}$; a detailed proof is given in Appendix~\ref{app:halfspace_cone}.
We use the duality rule for intersections of convex cones \cite{Rockafellar1970}: the dual of the intersection of finitely many convex cones equals the closure of the Minkowski sum of their dual cones,
\[
\biggl(\bigcap_{i=1}^m K_i\biggr)^* = \overline{\sum_{i=1}^m K_i^*}.
\]
If the intersection has nonempty interior, the closure may be omitted:
\[
\biggl(\bigcap_{i=1}^m K_i\biggr)^* = \sum_{i=1}^m K_i^*.
\]

Now take
\[\begin{array}{llllll}
\hat \rho_0 &= \sum_{a=1}^d (d-a+1)\ket{q_a}\bra{q_a}\\
&= \sum_{a=1}^d (d-a+1)\ket{\pi^{-1}(a)}{\pi^{-1}(a)}\\
&= \operatorname{diag}(d,\,d-1,\,d-2,\,\dots,\,2,\,1).
\end{array}\]
Clearly $\hat \rho_0$ is a positive definite matrix, $\hat \rho_0\in S_+^d$. Moreover,
\[\begin{aligned}
\mathrm{Tr}\big(F_a^{(\pi)} \hat \rho_0\big)
&= \bra{q_a}\hat \rho_0\ket{q_a} - \bra{q_{a+1}}\hat \rho_0\ket{q_{a+1}}\\
&=(d-a+1)-(d-a)=1>0,
\end{aligned}\]
i.e., $\hat \rho_0\in H_a$. As the strict feasibility holds, we have
\[
\mathcal{C}_\pi^* = S_+^d + \sum_{l=1}^{d-1} H_l^*
= \Big\{ Q + \sum_{l=1}^{d-1}\lambda_l F_l^{(\pi)} \;:\; Q\succeq0,\;\lambda_l\ge0\Big\},
\]
where $Q$ is any real symmetric positive-semidefinite matrix.
Therefore, $\Phi$ is a $\pi$-ORO if and only if $\sum_n K_n^\mathsf{T} K_n=\mathbb{I}$, $\langle i|K_n|j\rangle\in\mathbb{R}$, and for all $n,a$,
\begin{equation}\tag{R}
K_n^\mathsf{T} F_a^{(\pi)} K_n = Q_{n,a} + \sum_{l=1}^{d-1}\lambda_l F_l^{(\pi)},
\end{equation}
with $Q_{n,a}\succeq0$ and $\lambda_l\ge0$.

$(\Leftarrow)$: For any $\rho\in\mathcal{C}_\pi$,
\[
\mathrm{Tr}\big(K_n^\mathsf{T} F_a^{(\pi)} K_n \rho\big)
= \mathrm{Tr}(Q_{n,a}\rho) + \sum_{l=1}^{d-1} \lambda_l\,\mathrm{Tr}\big(F_l^{(\pi)}\rho\big).
\]
Since $Q_{n,a}\succeq0$ and $\rho\succeq0$, we have $\mathrm{Tr}(Q_{n,a}\rho)\ge0$.
For $\rho\in\mathcal{C}_\pi$, $\mathrm{Tr}(F_l^{(\pi)}\rho)\ge0$.
Combined with $\lambda_l\ge0$, this yields
$\mathrm{Tr}\big(K_n^\mathsf{T} F_a^{(\pi)} K_n \rho\big)\ge0$,
which implies $K_n \rho K_n^\mathsf{T}\in\mathcal{C}_\pi$. Hence $K_n \mathcal{C}_\pi K_n^\mathsf{T}\subseteq\mathcal{C}_\pi$, and $\Phi$ is a $\pi$-ORO.

\vspace{1em}
Recall $\widetilde K_n = P_\pi K_n P_\pi^\mathsf{T}$ and $A_a=\ket{a}\bra{a}-\ket{a+1}\bra{a+1}$.
By definition, $F_a^{(\pi)}=\ket{\pi^{-1}(a)}{\pi^{-1}(a)}-\ket{\pi^{-1}(a+1)}{\pi^{-1}(a+1)}$, and $P_\pi F_a^{(\pi)} P_\pi^\mathsf{T} = A_a$, $P_\pi^\mathsf{T} A_a P_\pi=F_a^{(\pi)}$.
Multiplying both sides of condition (R) by $P_\pi$ on the left and $P_\pi^\mathsf{T}$ on the right, we obtain
\[
P_\pi K_n^\mathsf{T} F_a^{(\pi)} K_n P_\pi^\mathsf{T}
= P_\pi Q_{n,a} P_\pi^\mathsf{T}
+ \sum_{l=1}^{d-1}\lambda_l\, P_\pi F_l^{(\pi)} P_\pi^\mathsf{T}.
\]
This simplifies to
\[
\widetilde K_n^\mathsf{T} A_a \widetilde K_n
= \widetilde Q_{n,a} + \sum_{l=1}^{d-1}\lambda_lA_l,
\]
namely,
\[
\widetilde K_n^\mathsf{T} A_a \widetilde K_n - \sum_{l=1}^{d-1}\lambda_l A_l = \widetilde Q_{n,a}\succeq0.
\]
This completes the proof.

\section{Proof of the Dual Cone of the Half-Space Cone}\label{app:halfspace_cone}
Recall the half-space cone
\(
H_a = \big\{ \rho \mid \operatorname{Tr}(F_a^{(\pi)} \rho)\ge0 \big\},
\)
and its dual cone is defined as
\(
H_a^*=\bigl\{\sigma \mid \operatorname{Tr}(\sigma \rho)\ge 0,\;\forall \rho\in H_a\bigr\}.
\)

$(\supseteq)$: Suppose $\sigma=\lambda F_a^{(\pi)}$ with $\lambda\ge 0$. For any $\rho\in H_a$,
\(
\operatorname{Tr}(\sigma\rho)
=\langle \sigma,\rho\rangle
=\lambda\,\bigl\langle F_a^{(\pi)},\rho\bigr\rangle\ge 0.
\)
Therefore $\bigl\{\lambda F_a^{(\pi)}: \lambda\ge 0\bigr\}\subseteq H_a^*$.

$(\subseteq)$: Take an arbitrary $\sigma\in H_a^*$. For any $\tau\in \bigl(F_a^{(\pi)}\bigr)^\perp$, we have
\(
\bigl\langle F_a^{(\pi)},\tau\bigr\rangle=0,
\)
which implies both $\tau\in H_a$ and $-\tau\in H_a$. Since $\sigma\in H_a^*$,
\(
\langle\sigma,\tau\rangle\ge 0,\quad \langle\sigma,-\tau\rangle\ge 0
\;\Longrightarrow\;
\langle\sigma,\tau\rangle=0,\quad \forall \tau\in \bigl(F_a^{(\pi)}\bigr)^\perp.
\)
It follows that
\(
\sigma\in \Bigl(\bigl(F_a^{(\pi)}\bigr)^\perp\Bigr)^\perp
=\operatorname{span}\bigl\{F_a^{(\pi)}\bigr\},
\)
so $\sigma=\lambda F_a^{(\pi)}$ for some real scalar $\lambda\in\mathbb{R}$.

Now set $\rho=F_a^{(\pi)}$. Observe that
\(
\bigl\langle F_a^{(\pi)},F_a^{(\pi)}\bigr\rangle
=\bigl\|F_a^{(\pi)}\bigr\|^2\ge 0,
\)
and hence $F_a^{(\pi)}\in H_a$. Substituting into the duality condition:
\[
\bigl\langle\sigma,F_a^{(\pi)}\bigr\rangle
=\lambda\bigl\langle F_a^{(\pi)},F_a^{(\pi)}\bigr\rangle
=\lambda\bigl\|F_a^{(\pi)}\bigr\|^2\ge 0.
\]
This forces $\lambda\ge 0$. Consequently,
\(
H_a^* \subseteq \bigl\{\lambda F_a^{(\pi)}: \lambda\ge 0\bigr\}.
\)
Combining the two inclusions, we conclude
\(
H_a^*=\bigl\{\lambda_a F_a^{(\pi)}: \lambda_a\ge 0\bigr\}.
\)

\section{Proof of Equation~(\ref{eq:relative entropy d=2})}\label{app:relative entropy}
Any state $\sigma\in\cR_\pi$ can be parametrized as
\[
\sigma=\frac12(I+s_x\sigma_x+s_z\sigma_z)=
\begin{pmatrix}
	\dfrac{1+s_z}{2}&\dfrac{s_x}{2}\\[2mm]
	\dfrac{s_x}{2}&\dfrac{1-s_z}{2}
\end{pmatrix},
\]
subject to $s_x^2+s_z^2\le1$. Therefore, the set of two-dimensional $\pi$-ordered real states corresponds to a disk on the $x$--$z$ plane inside the Bloch sphere.

For $d=2$, the two permutation cases yield
\[
\pi=\mathrm{id} \quad\Longleftrightarrow\quad \sigma_{11}\ge \sigma_{22} \quad\Longleftrightarrow\quad s_z\ge0,
\]
\[
\pi=(12) \quad\Longleftrightarrow\quad \sigma_{22}\ge \sigma_{11} \quad\Longleftrightarrow\quad s_z\le0.
\]
Thus $\mathcal R_{\mathrm{id}}$ is the upper half-disk, while $\mathcal R_{(12)}$ is the lower half-disk.

Recall the relative entropy $S(\rho\|\sigma) = -S(\rho)-\operatorname{Tr}(\rho\log\sigma)$. Since $S(\rho)$ is independent of $\sigma$, minimizing $S(\rho\|\sigma)$ over $\sigma\in\cR_\pi$ is equivalent to maximizing $\operatorname{Tr}(\rho\log\sigma)$.

Let
\(
t=\sqrt{s_x^2+s_z^2}, (s_x,s_z)=t(n_x,n_z),
\)
with $n_x^2+n_z^2=1$. Then
\(
\sigma=\frac12\big(I+t(n_x\sigma_x+n_z\sigma_z)\big),
\)
whose eigenvalues are $\lambda_\pm=\frac{1\pm t}{2}$. The matrix logarithm reads
\(
\log\sigma = \frac12\log\frac{1-t^2}{4}\,I + \operatorname{arctanh}(t)\,(n_x\sigma_x+n_z\sigma_z).
\)
Writing the Bloch components of $\rho$ as $r_x=\operatorname{Tr}(\rho\sigma_x)$, $r_y=\operatorname{Tr}(\rho\sigma_y)$, $r_z=\operatorname{Tr}(\rho\sigma_z)$, we obtain
\[
\operatorname{Tr}(\rho\log\sigma) = \frac12\log\frac{1-t^2}{4} + \operatorname{arctanh}(t)\,(r_xn_x+r_zn_z).
\]

We first optimize over $t$ for a fixed direction $(n_x,n_z)$. Define $k=r_xn_x+r_zn_z$ and consider
\[
f(t)= \frac12\log\frac{1-t^2}{4} +\operatorname{arctanh}(t)\,k.
\]
Differentiating:
\(
f'(t) = -\frac{t}{1-t^2} +\frac{k}{1-t^2} = \frac{k-t}{1-t^2}.
\)
The maximum occurs at $t=k$. It remains to maximize $k=r_xn_x+r_zn_z$ over admissible directions.

\textbf{1) Case $\pi=\mathrm{id}$ ($n_z\ge 0$).}
If $r_z\ge0$, the vector $(r_x,r_z)$ lies in the admissible upper half-plane. By the Cauchy–Schwarz inequality,
\[
r_xn_x+r_zn_z \le \sqrt{r_x^2+r_z^2}\sqrt{n_x^2+n_z^2} = \sqrt{r_x^2+r_z^2},
\]
with equality when $(n_x,n_z) = (r_x,r_z)/\sqrt{r_x^2+r_z^2}$. Hence $t=k_{\max}=\sqrt{r_x^2+r_z^2}$, which gives $s_x=r_x$, $s_z=r_z$. The maximizer is
\(
\sigma_*= \frac12(I+r_x\sigma_x+r_z\sigma_z) =\rho_{\mathrm R}.
\)

If $r_z<0$, the direction parallel to $(r_x,r_z)$ violates $n_z\ge0$. For any admissible $(n_x,n_z)$ with $n_z\ge0$, $r_zn_z\le0$, so
\[
k=r_xn_x+r_zn_z \le r_xn_x \le |r_x|,
\]
and equality is attained at $n_z=0$, $n_x=\operatorname{sgn}(r_x)$. Then $t=k_{\max}=|r_x|$, $s_x=r_x$, $s_z=0$, and the maximizer is the boundary state
\[
\sigma_\partial = \frac12(I+r_x\sigma_x) = \frac12\begin{pmatrix}
	1 & r_x\\
	r_x & 1
\end{pmatrix}.
\]

\textbf{2) Case $\pi=(12)$ ($n_z\le0$).}
If $r_z\le0$, the optimal state is $\rho_{\mathrm R}$; if $r_z>0$, the optimal state is $\sigma_\partial$.

Summarizing,
\[
\sigma_*=
\begin{cases}
	\rho_{\mathrm R},&\rho_{\mathrm R}\in\mathcal R_\pi,\\[1mm]
	\sigma_\partial,&\rho_{\mathrm R}\notin\mathcal R_\pi,
\end{cases}
\]
and therefore
\[
\J_S^\pi(\rho)=
\begin{cases}
	S(\rho\|\rho_{\mathrm R}) = S(\rho_{\mathrm R})-S(\rho), & \rho_{\mathrm R}\in\mathcal R_\pi,\\[1mm]
	S(\rho\|\sigma_\partial) = S(\sigma_\partial)-S(\rho), & \rho_{\mathrm R}\notin\mathcal R_\pi.
\end{cases}
\]

\section{Proof of Proposition~\ref{prop:pi-robust-2}}\label{app:robustness}
The derivation follows standard convex duality arguments; see, e.g., Ref.~\cite{Boyd2004}.
Recall the definition
\[
\J^\pi_R(\rho)= \min_{\tau\in\mathcal{D}(\mathcal{H})} \left\{ s\ge0 \;\bigg|\; \frac{\rho+s\tau}{1+s}\in\mathcal{R}_\pi \right\}.
\]
Set $\sigma=\frac{\rho+s\tau}{1+s}\in\mathcal{R}_\pi$. Rearranging gives
\(
\rho+s\tau=(1+s)\sigma \;\Longrightarrow\; (1+s)\sigma-\rho = s\tau \succeq 0.
\)
This yields an equivalent formulation:
\[
\J^\pi_R(\rho) = \min_{\sigma\in\mathcal{R}_\pi} \left\{ s\ge0 \;\bigg|\; (1+s)\sigma-\rho\succeq0 \right\}.
\]
Now introduce $\Gamma=(1+s)\sigma$. Since $\sigma\in\mathcal{R}_\pi$, we have $\Gamma\in\operatorname{cone}(\mathcal{R}_\pi)$. The constraints become
\[
\Gamma-\rho\succeq0,\qquad \operatorname{Tr}\Gamma = (1+s)\operatorname{Tr}\sigma = 1+s,
\]
so $s=\operatorname{Tr}\Gamma-1$. The primal SDP is
\[
\J^\pi_R(\rho) = \min_{\Gamma}\big\{\operatorname{Tr}\Gamma -1 \;\big|\; \Gamma-\rho \succeq 0,\ \Gamma\in\operatorname{cone}(\mathcal{R}_\pi)\big\}.
\]

The set $\mathcal{R}_\pi$ is defined by the ordering constraints $\operatorname{Tr}(D_k\Gamma)\ge0$, where
\[
D_k= |p_k\rangle\langle p_k| - |p_{k+1}\rangle\langle p_{k+1}|,\quad k=1,\ldots,d-1,\qquad p_k=\pi^{-1}(k).
\]
We introduce dual variables: $Y\succeq0$ for the matrix inequality $\Gamma-\rho\succeq0$, and $\lambda_k\ge0$ for the linear constraints $\operatorname{Tr}(D_k\Gamma)\ge0$. The Lagrangian reads
\[
\begin{aligned}
	L(\Gamma,Y,\lambda)
	&=\operatorname{Tr}\Gamma-1 -\operatorname{Tr}\big[Y(\Gamma-\rho)\big]-\sum_{k=1}^{d-1} \lambda_k\operatorname{Tr}(D_k\Gamma) \\
	&=\operatorname{Tr}(\rho Y)-1+\operatorname{Tr}\left[ \left( I-Y-\sum_{k=1}^{d-1}\lambda_kD_k \right)\Gamma \right].
\end{aligned}
\]
For $\inf_{\Gamma}L(\Gamma,Y,\lambda)$ to be finite, the coefficient of $\Gamma$ must vanish on the subspace of valid $\Gamma$, i.e.,
\[
I-Y-\sum_{k=1}^{d-1}\lambda_kD_k = 0.
\]
Under this condition, $L(\Gamma,Y,\lambda)=\operatorname{Tr}(\rho Y)-1$. If this condition fails, one may choose a direction of $\Gamma$ that drives the Lagrangian to $-\infty$, so such dual points are not feasible.

The dual feasible region is equivalent to $\operatorname{Tr}(\sigma Y)\le 1$ for all $\sigma\in\mathcal{R}_\pi$. Therefore the dual SDP is
\[
\J^\pi_R(\rho)= \max_{\substack{Y\succeq 0}}\big\{\operatorname{Tr}(\rho Y)-1 \;\big|\; \operatorname{Tr}(\sigma Y)\le 1,\ \forall\sigma\in\mathcal{R}_\pi\big\},
\]
which completes the proof.

\end{document}